\PassOptionsToPackage{table,dvipsnames}{xcolor}
\documentclass[onefignum,,onetabnum]{siamonline250211}

\crefname{section}{Section}{Sections}
\crefname{subsection}{Subsection}{Subsections}

\usepackage{array}
\usepackage{tabularx}
\usepackage{graphicx}
 \usepackage{booktabs}
 \usepackage{multirow}
\usepackage{diagbox}

\usepackage{bm}
\usepackage{dsfont}
\usepackage{subcaption}
\usepackage{xfrac}
\usepackage{subcaption}  

\newcommand{\revhl}[1]{#1}
\newcommand{\img}{\operatorname{img}}
\newcommand{\spanV}{\operatorname{span}}
\newcommand{\N}{\bm{\psi}}

\DeclareMathOperator*{\argmin}{argmin}
\DeclareMathOperator{\rank}{rank}

\usepackage{lipsum}
\usepackage{amsfonts}
\usepackage{graphicx}
\usepackage{epstopdf}
\usepackage{algorithmic}
\ifpdf
\DeclareGraphicsExtensions{.eps,.pdf,.jpg,.jpg}
\else
\DeclareGraphicsExtensions{.eps}
\fi

\usepackage{enumitem}
\setlist[enumerate]{leftmargin=.5in}
\setlist[itemize]{leftmargin=.5in}

\newsiamremark{remark}{Remark}
\newsiamremark{hypothesis}{Hypothesis}
\crefname{hypothesis}{Hypothesis}{Hypotheses}
\newsiamthm{claim}{Claim}

\headers{Optimizing Feature Types for Nonlinear Inpainting}{V. Chizhov, F. 
Jost, and J. Weickert}

\title{Optimizing Multiple Feature Types for Image Inpainting \\in the Linear and Nonlinear Setting\thanks{Accepted for publication in SIAM Journal on Imaging 
Sciences. The final Version of Record will be available through SIAM.\funding{This work has received funding from the European 
			Research Council (ERC) under the European Union's Horizon 2020 
			research and innovation programme (grant agreement no.\ 741215, ERC 
			Advanced Grant IN\-CO\-VID)}}}

\author{Vassillen Chizhov\thanks{Mathematical Image Analysis Group, Faculty of 
Mathematics and Computer Science, Saarland University, 66041 Saarbr\"ucken, 
Germany (\email{chizhov@cs.uni-saarland.de}, 
\email{ferdinanddjost@gmail.com}, \email{weickert@mia.uni-saarland.de})}
	\and Ferdinand Jost\footnotemark[2]
	\and Joachim Weickert\footnotemark[2]}

\usepackage{amsopn}
\DeclareMathOperator{\diag}{diag}

\begin{document}

\maketitle              

\begin{abstract}
Inpainting-based compression stores a carefully optimized subset of
the full image data and reconstructs the missing data by inpainting.
The quality of these lossy codecs depends decisively on the
stored data. So far, these data consist almost exclusively of
pixel locations along with their grayscale or color values. In the
present paper, we present a general theory and a practical framework
that allows to incorporate arbitrary features which can be described
by linear or nonlinear equations. This includes e.g.\ derivatives of
arbitrary order or local integrals. Our features can be combined
with linear or nonlinear inpainting operators. Moreover, we present
an algorithm that automatically optimizes the location and the type
of the selected feature. The approach of allowing different types of
optimized features turns inpainting-based compression into a more
general, versatile and powerful paradigm. Our experiments report a
consistent quality gain when increasing the number of feature types
from $1$ to $5$. With the same amount of stored data, the average
peak signal-to-noise improvement is $2.76$ dB for harmonic (homogeneous
diffusion) inpainting, and $1.82$ dB for edge-enhancing diffusion 
inpainting.  
\end{abstract}




\section{Introduction}
\label{sec:intro}

Inpainting-based compression methods are lossy codecs that have been 
introduced as an alternative to classical transform-based approaches 
such as JPEG~\cite{PM92} and JPEG2000~\cite{TM02}. They may offer
superior image quality if large compression rates are desired and
the amount of texture is not too high~\cite{GWWB08}.
The general idea is to store only a small and carefully selected 
subset of the data. In the decoding phase one reconstructs the 
missing data with an inpainting process \cite{BSCB00, GL14,  Sc15}.

\medskip 
Since the reconstruction quality depends crucially on the stored data, 
an appropriate data selection and optimization is of utmost importance. 
In contrast to transform-based codecs that use orthogonal or near 
orthogonal representations, where it is optimal to keep the 
coefficients with the largest absolute values, the data selection 
problem in inpainting-based cannot benefit from orthogonal representations
and becomes highly nontrivial: As is reviewed in Section 2, numerous 
methods have been proposed. However, most of them have
one thing in common: The feature type they store consists of optimized 
pixel locations and their function values (grayscale or color values).
A systematic exploration of different feature types and their joint
use is missing so far.
 
\medskip 
The goal of the present paper is to close this gap. We show the benefits 
of allowing a general set of feature types that can be characterized 
by linear or nonlinear equations. This may include for example 
derivative features of arbitrary order or local averages. 
We introduce an optimization strategy that simultaneously selects 
the location and the type of the feature. Our experiments demonstrate 
that this can lead to very substantial quality improvements without 
increasing the data budget.

In a recent conference paper~\cite{JCW23}, we have proposed a more 
general framework that allows arbitrary combinations of features 
expressed as linear equality constraints (e.g.~colors, derivatives, 
local integrals), along with an optimization strategy for selecting 
which features to store.

In the present work, we extend that framework to support nonlinear 
inpainting operators and nonlinear features, which can be 
formulated through nonlinear equality constraints. We further adapt 
and enhance the optimization algorithm from~\cite{JCW23} for this 
more general setting, leading to improved performance over our 
previous work and achieving qualitatively very good results in sparse 
image reconstruction.

	
\subsection{Our Contribution}
\label{sec:contribution}

In the current work we extend all of the results presented in 
our conference publication~\cite{JCW23}, that introduced linear 
features for linear inpainting arising from a minimization problem.
\begin{enumerate}
\item As far as we are aware our formulation, 
\eqref{eq:nlin_inpainting_nlc_semi_compact1} and 
\eqref{eq:nlin_inpainting_nlc_semi_compact2}, is the first 
framework for nonlinear inpainting with arbitrary nonlinear 
equality-constrained features. It generalizes 
the linear case discussed in~\cite{JCW23}, and is consistent with an 
optimization-based formulation when the inpainting operator corresponds 
to the gradient of an energy functional. 
Crucially, it also extends beyond this setting, accommodating inpainting 
processes that are not energy-derived -- most notably, edge-enhancing 
diffusion inpainting~\cite{GWWB08,We97}. We outline practical 
guidelines for numerically 
solving the proposed inpainting problem and demonstrate their 
effectiveness through a series of experiments.

\item We extend our spatial optimization strategy from~\cite{JCW23} 
to support nonlinear inpainting with nonlinear features 
(\cref{alg:voronoi-densification}). Our method 
automatically allocates the data budget across feature types and 
optimizes their spatial placement. 
Remarkably, this is achieved with a constant number of inpaintings, 
regardless of image resolution, even though the original search space 
grows exponentially with the number of pixels. 

\item We validate the robustness of our methods on a diverse set 
of natural images exhibiting a wide range of structures and 
frequencies. Our approach achieves results that surpass previous 
methods by a wide margin while using the same mask density. 
Moreover, owing to a better theoretical underpinning of 
our spatial optimization, we are able to improve upon our previous 
results from~\cite{JCW23} by up to $50$\% 
(see the last paragraph of \cref{sec:exp_harmonic}).
\end{enumerate}
Our work focuses exclusively on the inpainting 
and data optimization and as such we do not concern ourselves 
with coding aspects. 
\revhl{A thorough investigation of coding-related issues for feature-based 
representations would constitute a substantial research topic in its 
own right; see, for instance, the analysis in~\cite{MPW21}.}
Moreover, 
since the current work is already fairly extensive, 
we do not discuss the tonal optimization problem either -- for the 
linear setting we refer the reader to our conference paper~\cite{JCW23}.

\subsection{Outline}

We begin in \cref{sec:related} with a review of relevant related work.
\cref{sec:inpainting_operators} introduces the mathematical formulation 
of two prototypical inpainting operators -- one linear and one nonlinear -- 
which serve as reference points throughout the paper. 
In \cref{sec:linear_features_inpainting} we present our framework for 
linear and nonlinear inpainting with linear \revhl{equality constraint features}, 
along with solution strategies.
This is extended in \cref{eq:nonlinear_features_inpainting} to handle 
nonlinear \revhl{equality constraint features}, requiring more sophisticated 
methods, which we detail in \cref{sec:solution_strategies}. \cref{sec:spatial_optimization} 
describes our spatial optimization approach,
and \cref{sec:experiments} is devoted to experimental validation 
of our framework. We conclude 
the paper in \cref{sec:conclusions}, where we also briefly discuss future 
work.


\section{Related Work}
\label{sec:related}

In the current section we discuss works relevant to our feature 
inpainting, as well as to the spatial optimization method used to 
select the data to be stored.


\subsection{Feature Inpainting}

Some inpainting-based strategies allocate the gray or color value 
interpolation constraints at edges~\cite{Ca88,MBWF11} or 
isolines~\cite{SCSA04}, but they do not use different features 
than gray or color values. 
Approaches relying on a combination of grayscale data with 
discontinuities are presented in~\cite{HMWP13,JPW20,JPW21}. Unlike 
our methods, those use segmentation ideas which do not 
generalize to arbitrary features.

Other works have considered image reconstruction from features 
such as top points in scale-space \cite{KLDJ05}, zero-crossings \cite{RZ86}, 
junctions \cite{CCM97}, SIFT descriptors~\cite{DL18, WJP11, WZL19}, 
HOG descriptors~\cite{DL17,VKMT13},
local binary descriptors~\cite{DJAV14}, a number of 
shallow representations~\cite{DB16, MV14}, Bag-of-Visual-Words~\cite{KH14}, and positions and orientations~\cite{SSD25}.
KAZE features~\cite{ABD12} have also been proposed in the context 
of nonlinear scale spaces, however, the authors do not discuss reconstruction 
from such features.
While all of those works are theoretically interesting, none of them result in 
practical methods with a competitive quality for reconstructing natural 
images.

Methods that achieve better quality by considering gradient data 
are proposed in the works of Brinkmann et al.~\cite{BBG15} and 
Schneider et al.~\cite{SPHW16}. However they also do 
no generalize and do not allow for a combination 
of different features. A combination of local binary patterns 
and gray value interpolation is considered in the master's thesis of 
Sirazitdinov~\cite{Si19}, where a local binary pattern is 
coupled with each gray value.
Our conference publication~\cite{JCW23} provides a general and performant 
framework for harmonic inpainting with linear features. However, 
it has not been generalized to nonlinear inpainting and nonlinear 
features. In the current work we provide such an extension.


\subsection{Spatial Optimization}

The choice of which data to store -- known as spatial optimization -- is 
critical for achieving high reconstruction quality. However, this task 
poses a challenging combinatorial problem, often requiring carefully 
designed greedy algorithms to find effective solutions within a 
reasonable time. 
Let us now briefly review the main categories of approaches that have 
been explored.


\paragraph{Analytic Approaches}

Belhachmi et al.~\cite{BBBW09} have derived a spatial data selection 
framework for harmonic inpainting in the continuous setting. It is
based on the theory of shape optimization. To transfer these results
to the practically relevant discrete setting, error diffusion approaches 
are used. While the theory gives real optimality results as long as one
stays within the continuous setting and it also allows for very fast 
algorithms without requiring explicit inpaintings, the quality of the 
final result is compromised by the inherent limitations of the 
halftoning step. Moreover, so far the theory is only available for
harmonic inpainting.


\paragraph{Nonsmooth Optimization Strategies}

By relaxing the spatial optimization problem from a discrete 
combinatorial problem to a continuous one, one can apply (sub-)gradient 
descent methods, primal-dual solvers, and optimal 
control ideas~\cite{BLPP17,CRP14,HSW13,OCBP14}. 
Such strategies require a projection step at the end to go back 
to the discrete formulation which reduces the quality~\cite{HW15_2}. 
Nevertheless, they achieve state-of-the-art results but are 
generally prohibitively expensive. We consider greedier and more 
efficient strategies, but we note that developing a gradient-based method 
for optimizing different features is certainly an interesting avenue 
for future work.


\paragraph{Sparsification Methods}

While the combinatorial problem of finding the globally optimal mask 
pixels in a digital image is practically very hard (if not intractable) 
to solve exactly, the sparsification approach of Mainberger et 
al.~\cite{MHWT12} constitutes a greedy heuristics to find a reasonable 
local optimum. Starting with a full mask, one randomly nominates a
specified percentage of all mask pixels for removal, computes the local 
error after their testwise removal, and accepts only those removals where
this error magnitude is small. This idea is applied iteratively until 
the desired target density is reached. Sparsification methods are 
popular due to their simplicity and wide applicability to all types of 
inpainting operators.  However, their quality suffers from the local 
error measure which is computed only in the mask pixel. Moreover, they 
are relatively slow since they may require many inpaintings.


\paragraph{Densification Methods}

Densification approaches proceed in the opposite way as sparsification
methods: They start with an empty mask and gradually add mask pixels
until the target density is reached. One class uses constrained data 
structures such as subdivision trees~\cite{DNV97,GWWB08,PHNH16,SPME14},
which supports efficient encoding, but may require sophisticated 
inpainting operators to achieve high quality in spite of such 
suboptimal masks. More general unconstrained settings go back to 
Karos et al.~\cite{KBPW18}, who introduce error maps
to find good pixel locations for exemplar-based inpainting.
Variants of these ideas have been applied to diffusion-based 
inpainting~\cite{CW21,DAW21,JCW23,TPM20} and Delaunay-based linear 
spline inpainting~\cite{EA17}. With their more area-based way of
measuring the impact of a mask pixel on the inpainting result and
their coarse-to-fine strategy that they share with many successful
optimization methods, modern densification approaches can be 
qualitatively superior to the purely local error measurements and
the fine-to-coarse progress in sparsification approaches. 
They may also be more efficient by requiring less inpainting steps to 
reach the (usually relatively small) target density. In practice, this
class constitutes one of the most attractive data selection paradigms.

Our method builds on the densification approach proposed in our
previous work~\cite{JCW23}, and, like in~\cite{CW21, DAW21}, it
relies on an adaptive partition of the domain in order to guide
the selection of stored data and avoid clustering.


\paragraph{Relocation Approaches}

A class of methods that resemble simulated annealing ideas are 
techniques that allow probabilistically swapping out some stored data 
for new one. Since such methods make very weak assumptions about the 
inpainting operator they are very general and allow escaping from 
local minima. Mainberger et al.~\cite{MHWT12} have devised such 
a strategy specifically in the context of inpainting. While their 
approach allows to reduce the mean squared error (MSE), the 
computational cost is considerable. 
The above method is directly applicable to any of the discussed 
inpaintings in our work. It can be applied as a post-processing 
step after our spatial optimization if runtime is not a major concern. 
We also note that a specialized method with similar global relocation 
ideas has been developed for linear spline interpolation on 
triangulations~\cite{MMCB18}. 


\paragraph{Neural Approaches}

In recent years, a number of reasonably well-performing neural-based 
approaches have been advocated for spatial optimization. Dai et 
al.~\cite{DCPCWK19} have developed a deep learning method for 
adaptive sampling in the context of inpainting. Peter~\cite{P23} 
used Wasserstein GANs for a joint training of an inpainting and 
spatial optimization. For harmonic inpainting, Alt et al.~\cite{APW22} 
have proposed a network for spatial optimization, with its efficiency 
improved further by Peter et al.~\cite{PSAW22}. In order to make this 
applicable to 4K images, Schrader et al.~\cite{SPKW23} have devised 
a coarse-to-fine approach that splits the image into patches.
However, as none of the networks consider different features, they are 
not directly applicable to our problem. Nevertheless, developing a 
neural-based spatial optimization for our feature inpainting is an 
appealing direction for future research.


\section{Inpainting Operators}
\label{sec:inpainting_operators}

In the current work we consider harmonic inpainting (also known as homogeneous 
diffusion inpainting) as a classical representative of a linear inpainting 
operator that can benefit substantially from careful data 
optimization~\cite{PHNH16}, and 
edge-enhancing diffusion (EED) inpainting as a representative of a 
nonlinear operator that offers some of the best results in inpainting-based 
image compression~\cite{SPME14}. 
While our theory is general and covers virtually any differentiable 
inpainting operator which fills in missing information by solving a linear or 
nonlinear system of equations of the form $\N(\bm{u}) = \bm{0}$, the above 
serve as adequate illustrations of the linear and nonlinear case. 
We discuss the continuous and discrete formulation of those in 
\Cref{sec:harmonic_inpainting} and 
\Cref{sec:EED_inpainting}.


\subsection{Homogeneous Diffusion Inpainting}
\label{sec:harmonic_inpainting}

\begin{figure}
    \begin{tabular}{ccc}
    \includegraphics[width=0.3\textwidth]{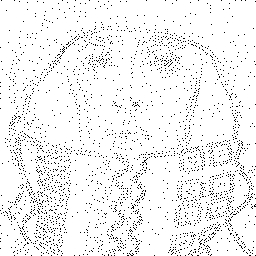} &
    \includegraphics[width=0.3\textwidth]{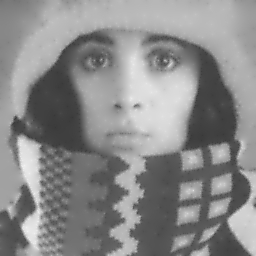} &
    \includegraphics[width=0.3\textwidth]{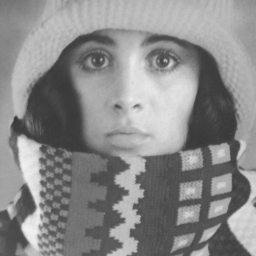} \\
   mask $\bm{c}$ & harmonic inpainting $\bm{u}$ & original image 
   \emph{trui} $\bm{f}$
    \end{tabular}
    \caption{Illustration of harmonic inpainting with a $5\%$ mask. Inpainting MSE: $38.24$.}
    \label{fig:trui_harmonic}
\end{figure}

We briefly review the idea behind harmonic or homogeneous diffusion 
inpainting~\cite{Ca88}. Although it is one of the simplest linear 
inpainting operators, it can perform surprisingly well in practice, 
if the data are chosen carefully. 
We assume that we wish to reconstruct a 
function $u:\Omega\to\mathbb{R}$ given known data $f|_{K}$ over some 
set $K\subset\Omega$, where $\Omega$ is the image domain, and $K$ 
are the parts of it where we store data.
To carry out the reconstruction we 
require that $u$ is harmonic over $\Omega\setminus K$, and 
interpolating $f$ on $K$. Additionally, reflecting boundary 
conditions are taken on $\partial\Omega$
\begin{alignat}{3}
    -\Delta u(\bm{x}) &= 0, &\quad& \bm{x}\in \Omega\setminus K, \\
    u(\bm{x}) &= f(\bm{x}), &\quad& \bm{x}\in K, \\
    \partial_{\bm{n}} u(\bm{x}) &= 0, &\quad& 
    \bm{x}\in\partial\Omega,
\end{alignat}
where $\bm{n}$ is the normal to $\partial\Omega$. By introducing the 
indicator function 
\begin{equation}
    c(\bm{x}) = \mathds{1}_{K}(\bm{x}) = \begin{cases}
        1, &\text{for } \bm{x}\in K, \\
        0, &\text{for } \bm{x}\not\in K,
    \end{cases}
\end{equation}
we can rewrite the above as follows:
\revhl{
\begin{alignat}{3}
    (1-c(\bm{x}))(-\Delta)u(\bm{x}) &= 0, 
    &\quad& \bm{x}\in \Omega, \\
    c(\bm{x})u(\bm{x}) &= c(\bm{x})f(\bm{x}), 
    &\quad& \bm{x}\in\Omega, \\
    \partial_{\bm{n}} u(\bm{x}) &= 0, &\quad& 
    \bm{x}\in\partial\Omega.
\end{alignat}}
Combining the first and second equation yields
\begin{alignat}{3}
    \label{eq:harmonic_inp_cont_compact}
    \bigl(c(\bm{x})+(1-c(\bm{x}))(-\Delta)\bigr)u(\bm{x}) &= 
    c(\bm{x})f(\bm{x}), 
    &\quad& \bm{x}\in \Omega, \\
    \partial_{\bm{n}} u(\bm{x}) &= 0, &\quad& 
    \bm{x}\in\partial\Omega.
\end{alignat}
In image processing $\partial\Omega$ is usually the image's bounding 
rectangle and $\Delta$ is discretised with the standard $5$-point 
finite difference stencil, while the homogeneous Neumann boundary 
conditions are implemented by reflecting a single layer of pixels 
at the boundary. 
In the discrete setting $\bm{u},\,\bm{f}\in\mathbb{R}^N$ are 
$N$-dimensional vectors, where $N$ denotes the number of image pixels. 
The indicator function $c$ then becomes the binary \emph{mask} vector 
$\bm{c}\in\{0,1\}^N$ with corresponding \emph{mask matrix} 
$\bm{C} = \diag(\bm{c})\in\{0,1\}^{N\times N}$. Let 
$\bm{L}\in\mathbb{R}^{N\times N}$ be the finite difference matrix 
approximating $-\Delta$ with reflecting boundary conditions. 
Then the discrete counterpart of the reconstruction problem 
is given as
\begin{align}
    \label{eq:lin_inpainting_semi_compact1}
    (\bm{I}-\bm{C})\bm{L}\bm{u} &= \bm{0}, \\
    \label{eq:lin_inpainting_semi_compact2}
    \bm{C}\bm{u} &= \bm{C}\bm{f},
\end{align}
or more compactly
\begin{equation}
    \label{eq:discrete_harmonic_inpainting}
    (\bm{C}+(\bm{I}-\bm{C})\bm{L})\bm{u} = \bm{C}\bm{f}.
\end{equation}
This problem has a unique solution provided the mask is 
non-empty~\cite{MBWF11}. An illustration of the inpainting can be seen 
in \Cref{fig:trui_harmonic}.
A straightforward generalisation, spanning a whole family of inpainting 
operators, is $\alpha$-harmonic inpainting~\cite{AWA19}
\begin{equation}
    (\bm{C}+(\bm{I}-\bm{C})\bm{L}^{\alpha})\bm{u} = \bm{C}\bm{f}.
\end{equation}
In the current work, instead of focusing on various extensions 
of the inpainting operator, we study extensions of the types of data 
being interpolated. It turns out that the latter can in fact modify 
the operator itself (\cref{sec:compression_inpainting_operators}) 
in some sense, such that it becomes potentially much more powerful.


\subsection{Edge-Enhancing Diffusion Inpainting}
\label{sec:EED_inpainting}

As a practically relevant example for the nonlinear setting we consider 
edge-enhancing diffusion (EED) inpainting. The latter has been used to 
achieve very good results for image reconstruction from 
sparse data~\cite{GWWB08,PHNH16, SPME14}. 
The boundary value problem that gives the desired reconstruction 
is as follows:
\revhl{
\begin{alignat}{3}
    -\bm{\nabla}^{\top}(\bm{D}(\bm{\nabla}u_{\sigma})\bm{\nabla} u) &= 0, 
    &\quad& \text{ on } \Omega\setminus K, \\
    u &= f, &\quad& \text{ on } K, \\
    \partial_{\bm{n}} u &= 0, &\quad& 
    \text{ on }\partial\Omega.
\end{alignat}}
Here $u_{\sigma}$ is a Gaussian-smoothed version of $u$, 
$u_{\sigma} = K_{\sigma} * u$, where $K_{\sigma}$ is a Gaussian with 
standard deviation $\sigma$ and the convolution satisfies reflecting 
boundary conditions on $\partial\Omega$. \revhl{The diffusion tensor 
$\bm{D}(\bm{\nabla}u_{\sigma})$ is derived from the structure tensor.
Let 
\begin{equation}
\bm{J} = \bm{\nabla}u_{\sigma}\bm{\nabla}u_{\sigma}^{\top}
=
\bm{S} \operatorname{\mathbf{diag}}(\mu_{\min},\mu_{\max}) \bm{S}^{\top}
\end{equation}
be the eigendecomposition of the structure tensor. The diffusion
 tensor $\bm{D}$ is then defined by 
 \begin{equation}
 \bm{D} = \bm{S} \operatorname{\mathbf{diag}}(\mu'_{\min},\mu'_{\max}) \bm{S}^{\top},
 \end{equation}
 with $\mu'_{\min}=1$, $\mu'_{\max} = \frac{1}{\sqrt{1+\mu_{\max}/\lambda^2}}$. 
That is, the smaller eigenvalue is replaced by $1$, while the Charbonnier diffusivity 
with contrast parameter $\lambda>0$ is applied to the larger eigenvalue.}

\begin{figure}
    \begin{tabular}{ccc}
    \includegraphics[width=0.3\textwidth]{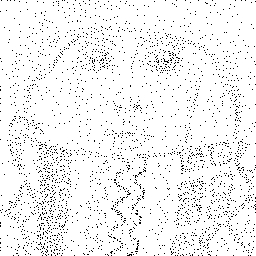} &
    \includegraphics[width=0.3\textwidth]{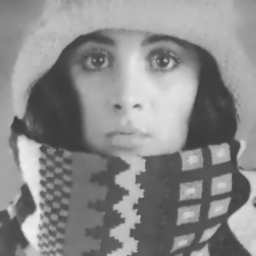} &
    \includegraphics[width=0.3\textwidth]{resources/trui/trui.jpg} \\
   mask $\bm{c}$ & EED inpainting $\bm{u}$ & original image 
   \emph{trui} $\bm{f}$
    \end{tabular}
    \caption{Illustration of EED inpainting with a $5\%$ mask. 
    Inpainting MSE: $14.95$.}
    \label{fig:trui_EED}
\end{figure}

We use the discretisation of anisotropic diffusion from~\cite{WWW13} 
which results in the quasi-linear system of equations
\begin{equation}
    \label{eq:discrete_EED_inpainting}
    (\bm{C}+(\bm{I}-\bm{C})\bm{L}(\bm{u}))\bm{u} = \bm{C}\bm{f}.
\end{equation}
Note that quasi-linearity means that the system is still nonlinear, 
however, the nonlinearity takes the special form 
$\N(\bm{u}) = \bm{L}(\bm{u})\bm{u}$.
An illustration of the inpainting can be seen in \Cref{fig:trui_EED}.
A more generic nonlinear operator $\N:\mathbb{R}^N\to\mathbb{R}^N$ 
(not necessarily quasi-linear) would result in the following formulation
\begin{equation}
    \label{eq:nonlinear_inpainting_formulation_compact}
    \bm{C}\bm{u}+(\bm{I}-\bm{C})\N(\bm{u}) = \bm{C}\bm{f}.
\end{equation}
%


\section{Linear Feature Inpainting}
\label{sec:linear_features_inpainting}

In our conference paper~\cite{JCW23} we have introduced a framework for 
augmenting harmonic inpainting to be able to handle any features 
that can be written through linear equality constraints. We briefly review 
the key ideas behind this in 
\cref{sec:linear_constrained_harmonic_inpainting_minimization}, and throughout 
the rest of \cref{sec:linear_features_inpainting} we extend the framework to 
be able to handle nonlinear inpainting operators with linear 
\revhl{equality constraint features}.

In \cref{sec:compression_inpainting_operators} we make explicit the fact 
that the linear equality constraints induce a projection of the 
inpainting operator on the kernel of the constraints. In 
\cref{sec:linear_constrained_linear_inpainting} we relate this to 
the classical inpainting formation from \eqref{eq:discrete_harmonic_inpainting}. 
We also emphasize that despite of the optimization viewpoint adopted 
in~\cite{JCW23} and \cref{sec:compression_inpainting_operators}, the inpainting 
operator does not need to arise as the gradient of some energy. This is crucial, 
as for instance powerful inpainting operators such as EED  
do not have an energy~\cite{We21}. 

In \cref{sec:linear_constrained_nonlinear_inpainting} we extend the framework 
to nonlinear inpainting operators, and in 
\cref{sec:linear_constrained_quasilinear_inpainting} we specialize this 
to quasi-linear inpainting operators such as EED.


\subsection{Linear-Constrained Harmonic Inpainting}
\label{sec:linear_constrained_harmonic_inpainting_minimization}

The discrete harmonic inpainting formulation 
in \Cref{eq:discrete_harmonic_inpainting} can at most interpolate pixel-wise 
color data. This is already quite powerful in practice, but nevertheless, 
the question arises whether one can use more complex 
features. In our prior work~\cite{JCW23} we reformulate 
the harmonic inpainting problem as a constrained optimization problem:
\begin{equation}
\label{eq:continuous_variational_formulation}
\begin{gathered}
    \min_u\frac{1}{2}\int_{\Omega}\|\bm{\nabla} u(\bm{x})\|^2\,d\bm{x} 
    = \min_u \frac{1}{2}\int_{\Omega}u(\bm{x})(-\Delta) u(\bm{x})\,d\bm{x},\\ 
    \textrm{such that} \,\, u(\bm{x}) = f(\bm{x}), \,\, \bm{x}\in K.
\end{gathered}
\end{equation}
Its discrete counterpart is then given as
\begin{equation}
    \min_{\bm{C}\bm{u}=\bm{C}\bm{f}} \frac{1}{2}\|\bm{D}\bm{u}\|^2_2 = 
    \min_{\bm{C}\bm{u}=\bm{C}\bm{f}}\frac{1}{2}\bm{u}^{\top}\bm{L}\bm{u}.
\end{equation}
In the above formulation it is straightforward to replace the 
mask matrix $\bm{C}$, in the constraint $\bm{C}\bm{u}=\bm{C}\bm{f}$, 
with any other matrix $\bm{A}\in\mathbb{R}^{M\times N}$.
As demonstrated in~\cite{JCW23}, this results in a practical approach 
for implementing interpolation of any combination of features that 
can be represented through linear equality constraints. 
Moreover, the inpainting problem remains 
linear. It is made unconstrained by using Lagrange multipliers:
\begin{equation}
    \min_{\bm{A}\bm{u}=\bm{A}\bm{f}}\frac{1}{2}\bm{u}^{\top}\bm{L}\bm{u} 
    \implies 
    \min_{\bm{u}\in\mathbb{R}^N}\max_{\bm{\lambda}\in\mathbb{R}^M}
    \frac{1}{2}\bm{u}^{\top}\bm{L}\bm{u} 
    + \bm{\lambda}^{\top}\bm{A}(\bm{u}-\bm{f}).
\end{equation}
The solution then satisfies the symmetric but indefinite saddle 
point system
\begin{equation}
    \label{eq:augmented_system_harmonic}
    \begin{bmatrix}
        \frac{1}{2}(\bm{L}+\bm{L}^{\top}) & \bm{A}^{\top} \\
        \bm{A} & \bm{0}
    \end{bmatrix} \begin{bmatrix}
        \bm{u} \\ \bm{\lambda}
    \end{bmatrix}
    =
    \begin{bmatrix}
        \bm{0} \\ 
        \bm{A}\bm{f}
    \end{bmatrix}.
\end{equation}
The solution is unique whenever the projection  
of $\frac{1}{2}(\bm{L}+\bm{L}^{\top})$ onto the kernel of $\bm{A}$ is 
non-singular. For instance the Laplacian matrix $\bm{L}$ has 
a kernel spanned by the vector $\bm{1}$, then if 
$\bm{A}\cdot\bm{1}\ne \bm{0}$ we get that the system matrix 
in \eqref{eq:augmented_system_harmonic} is non-singular. Consequently, 
for the Laplacian it suffices to include an average value constraint 
in $\bm{A}$ or a single value interpolation constraint, in order to 
guarantee uniqueness of the solution.

We note that for the Laplacian and 
other discretisations that result in symmetric matrices we have 
$\frac{1}{2}(\bm{L}+\bm{L}^{\top}) = \bm{L}$, however, this is not 
true if $\bm{L}$ is non-symmetric. It turns out that one can be 
more general, which also allows for the support of 
operators that are not the result of a minimization problem. 
We discuss this in the next subsection.

\subsection{Projection of Inpainting Operators}
\label{sec:compression_inpainting_operators}

We could have approached the linearly constrained minimization problem 
\begin{equation}
    \min_{\bm{A}\bm{u}=\bm{A}\bm{f}}\frac{1}{2}\bm{u}^{\top}\bm{L}\bm{u} 
\end{equation}
in a slightly different manner. Let the columns of 
$\bm{V}_{\ker} \in\mathbb{R}^{N\times (N-r)}$ form an orthonormal basis 
for the kernel of $\bm{A}$, and the columns of 
$\bm{V}_{\img}\in\mathbb{R}^{N\times r}$ form an orthonormal basis for 
its orthogonal complement (which has 
the same span as $\bm{A}^{\top}$). Then we may decompose $\bm{u}$ 
as a sum of vectors from the two subspaces as follows:
\begin{equation}
    \bm{u} = \bm{u}_{\img} + \bm{u}_{\ker} = 
    \bm{V}_{\img} \bm{s}_{\img} + \bm{V}_{\ker}\bm{s}_{\ker}.
\end{equation}
Plugging this into the constraints yields
\begin{equation}
\bm{A}\bm{u} = 
\bm{A}(\bm{V}_{\img} \bm{s}_{\img} + \bm{V}_{\ker}\bm{s}_{\ker}) = 
\bm{A}\bm{V}_{\img}\bm{s}_{\img} = \bm{A}\bm{f} \implies 
\bm{s}_{\img} = \bm{V}_{\img}^{\top}\bm{f}.
\end{equation}
Similarly, plugging the decomposition into the objective function 
results in
\begin{equation}
\begin{aligned}
    E(\bm{u}) &= \frac{1}{2}\bm{u}^{\top}\bm{L}\bm{u} 
    = \frac{1}{2}(\bm{V}_{\img}\bm{s}_{\img}
    +\bm{V}_{\ker}\bm{s}_{\ker})^{\top}\bm{L}(\bm{V}_{\img}\bm{s}_{\img}
    +\bm{V}_{\ker}\bm{s}_{\ker}) \\
    &= \frac{1}{2}\bm{s}_{\ker}^{\top}\bm{V}_{\ker}^{\top}\bm{L}
    \bm{V}_{\ker}\bm{s}_{\ker} +
    \bm{s}_{\ker}^{\top}\bm{V}_{\ker}^{\top}\frac{\bm{L}+\bm{L}^{\top}}{2}
    \bm{V}_{\img}\bm{s}_{\img}
    +
    \frac{1}{2}\bm{s}_{\img}^{\top}\bm{V}_{\img}^{\top}\bm{L}
    \bm{V}_{\img}\bm{s}_{\img}.
\end{aligned}
\end{equation}
We can now use $\bm{s}_{\img} = \bm{V}_{\img}^{\top}\bm{f}$ 
and minimize the energy only w.r.t.\ $\bm{s}_{\ker}$, 
$\bm{\nabla}_{\bm{s}_{\ker}}E(\bm{s}) = \bm{0}$, which 
yields the following linear system:
\begin{equation}
    \bm{V}_{\ker}^{\top}\frac{\bm{L}+\bm{L}^{\top}}{2}\bm{V}_{\ker}\bm{s}_{\ker} 
    = -\bm{V}_{\ker}^{\top}\frac{\bm{L}+\bm{L}^{\top}}{2}
    \bm{V}_{\img}\bm{s}_{\img} = 
    -\bm{V}_{\ker}^{\top}\frac{\bm{L}+\bm{L}^{\top}}{2}
    \bm{V}_{\img}\bm{V}_{\img}^{\top}\bm{f}.
\end{equation}
It is now clear that the above has a unique solution only when the 
projection 
$\bm{V}_{\ker}^{\top}\frac{\bm{L}+\bm{L}^{\top}}{2}\bm{V}_{\ker}$ of 
$\frac{\bm{L}+\bm{L}^{\top}}{2}$ on the kernel 
$\ker(\bm{A}) = \spanV(\bm{V}_{\ker})$ of $\bm{A}$ is non-singular. 
While we derived this for the minimization case, the projection  
formulation is more general, and we could have picked 
some arbitrary linear operator $\bm{L}\in\mathbb{R}^{N\times N}$, 
not necessarily positive semi-definite or symmetric, and considered instead
\begin{equation}
    \label{eq:inpainting_operator_compression}
    \bm{V}_{\ker}^{\top}\bm{L}\bm{V}_{\ker}\bm{s}_{\ker}  = 
    -\bm{V}_{\ker}^{\top}\bm{L}
    \bm{V}_{\img}\bm{V}_{\img}^{\top}\bm{f}.
\end{equation}
\Cref{eq:inpainting_operator_compression} is of course 
equivalent to the augmented system
\begin{equation}
    \label{eq:lin_lin_aug}
    \begin{bmatrix}
        \bm{L} & \bm{A}^{\top} \\
        \bm{A} & \bm{0}
    \end{bmatrix} \begin{bmatrix}
        \bm{u} \\ \bm{\lambda}
    \end{bmatrix}
    =
    \begin{bmatrix}
        \bm{0} \\ 
        \bm{A}\bm{f}
    \end{bmatrix},
\end{equation}
but it makes explicit the fact that we are really solving a problem 
for a projection of $\bm{L}$ onto the kernel of the constraints. 
This formulation also hints at the fact that this may be the case 
in the setting of a nonlinear inpainting 
(\cref{sec:linear_constrained_nonlinear_inpainting}), and 
also in the case of nonlinear constraints, if one 
considers their linearization 
(\cref{sec:nonlinear_constrained_nonlinear_inpainting}).

\subsection{Linear-Constrained Linear Inpainting}
\label{sec:linear_constrained_linear_inpainting}

\revhl{In order to relate \eqref{eq:inpainting_operator_compression} to 
the classical formulation $(\bm{C}+(\bm{I}-\bm{C})\bm{L})\bm{u} = \bm{C}\bm{f}$ 
from \eqref{eq:discrete_harmonic_inpainting}, 
we denote $\bm{P} =\bm{V}_{\img}\bm{V}_{\img}^{\top}$ 
and proceed as follows:
\begin{align}
        \bm{V}_{\ker}^{\top}\bm{L}\bm{V}_{\ker}\bm{s}_{\ker}  &= 
    -\bm{V}_{\ker}^{\top}\bm{L}
    \bm{V}_{\img}\bm{V}_{\img}^{\top}\bm{f} \\
    \bm{V}_{\ker}\bm{V}_{\ker}^{\top}\bm{L}\bm{V}_{\ker}
    \bm{V}_{\ker}^{\top}\bm{u}
    &= -\bm{V}_{\ker}\bm{V}_{\ker}^{\top}\bm{L}
    \bm{V}_{\img}\bm{V}_{\img}^{\top}\bm{f} \\
    (\bm{I}-\bm{P})\bm{L}(\bm{I}-\bm{P})\bm{u}&=
    -(\bm{I}-\bm{P})\bm{L}\bm{P}\bm{f} \\
    (\bm{I}-\bm{P})\bm{L}(\bm{I}-\bm{P})\bm{u}&=
    -(\bm{I}-\bm{P})\bm{L}\bm{P}\bm{u} \\
    \bm{P}\bm{u}+ (\bm{I}-\bm{P})\bm{L}\bm{u}&= \bm{P}\bm{u} \\
    (\bm{P}+ (\bm{I}-\bm{P})\bm{L})\bm{u} &= \bm{P}\bm{f}.
\end{align}
In the above derivation, the first line is simply \eqref{eq:inpainting_operator_compression}. 
To obtain the second line, we multiply both sides by $\bm{V}_{\ker}$ and use 
$\bm{s}_{\ker}=\bm{V}_{\ker}^{\top}\bm{u}$. For the third line, we use 
$\bm{P}=\bm{V}_{\img}\bm{V}_{\img}^{\top}$ and the identity 
$\bm{V}_{\ker}\bm{V}_{\ker}^{\top}=\bm{I}-\bm{P}$. In the fourth line, we use 
$\bm{P}\bm{f}=\bm{P}\bm{u}$, which follows from 
$\bm{V}_{\img}^{\top}\bm{f}=\bm{s}_{\img}$. To obtain the fifth line, 
we move the right-hand side term to the left-hand side, combine the 
two into $(\bm{I}-\bm{P})\bm{L}\bm{u}$, and then add 
$\bm{P}\bm{u}$ to both sides. Finally, we again use 
$\bm{P}\bm{u}=\bm{P}\bm{f}$ to obtain the last line.}

\begin{figure}
    \begin{tabular}{ccc}
    \includegraphics[width=0.3\textwidth]{resources/trui/trui_c1_d0_05_EED0_ni30_ns1.jpg} &
    \includegraphics[width=0.3\textwidth]{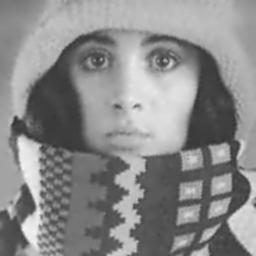} &
    \includegraphics[width=0.3\textwidth]{resources/trui/trui.jpg} \\
   gray values, MSE: $38.24$ & five features, MSE: $14.89$ & original image 
   \emph{trui} $\bm{f}$
    \end{tabular}
    \caption{Harmonic inpainting using only gray-value features versus five features (gray values, $\partial_x$, $\partial_y$, $3\times 3$ and $5\times 5$ binomial kernels). The mask density is $5\%$ in both cases. \textbf{The reconstruction error is more than halved despite using the same inpainting operator and mask density}. Note the significantly improved texture reconstruction of the hat.}
    \label{fig:trui_harmonic_feat_comp}
\end{figure}

Thus, we see that the inpainting problem with generalized features 
requires replacing the mask matrix $\bm{C}$ with the orthogonal 
projection matrix $\bm{P}$ that projects onto 
$\spanV(\bm{A}^{\top})$:
\begin{equation}
    (\bm{C}+(\bm{I}-\bm{C})\bm{L})\bm{u} = \bm{C}\bm{f}\implies 
    (\bm{P}+ (\bm{I}-\bm{P})\bm{L})\bm{u} = \bm{P}\bm{f}, 
\end{equation}
where the projection matrix satisfies:
\begin{equation}
    \bm{P}^2=\bm{P}, \,\, \bm{P}^{\top}=\bm{P},\,\, 
    \spanV(\bm{P}) = \spanV(\bm{A}^{\top}), \,\, 
    \ker(\bm{P}) = \ker(\bm{A}).
\end{equation}
The orthogonal projection can be formulated in terms of any 
$(1,4)$-inverse\footnote{
\revhl{$\bm{M}^{(1,4)}$ is called a $(1,4)$-inverse of $\bm{M}$ if it 
satisfies the 1st and 4th Moore-Penrose conditions:}  
$\bm{M}\bm{M}^{(1,4)}\bm{M} = \bm{M}$ and $(\bm{M}^{(1,4)}\bm{M})^{\top} 
= \bm{M}^{(1,4)}\bm{M}$.}
of $\bm{A}$ as $\bm{P} = \bm{A}^{(1,4)}\bm{A}$. We may also write
\begin{align}
    \label{eq:lin_inpainting_lc_semi_compact1}
    (\bm{I}-\bm{P})\bm{L}\bm{u} &= \bm{0} \\
    \label{eq:lin_inpainting_lc_semi_compact2}
    \bm{A}\bm{u} &= \bm{A}\bm{f},
\end{align}
similarly to \eqref{eq:lin_inpainting_semi_compact1} and 
\eqref{eq:lin_inpainting_semi_compact2}.
This formulation is of course also equivalent to the 
augmented system formulation from \eqref{eq:lin_lin_aug}
\begin{equation}
    \begin{bmatrix}
        \bm{L} & \bm{A}^{\top} \\
        \bm{A} & \bm{0}
    \end{bmatrix}
    \begin{bmatrix}
        \bm{u} \\ \bm{\lambda}
    \end{bmatrix}
    = \begin{bmatrix}
        \bm{0} \\ \bm{A}\bm{f}
    \end{bmatrix},
\end{equation}
because $\bm{\lambda}$ can compensate for any part of $\bm{L}\bm{u}$ in the 
span of $\bm{A}^{\top}$.
The augmented system~\eqref{eq:lin_lin_aug} is especially appealing when 
computing a matrix-vector product with $\bm{P}$ is expensive. 
If $\bm{L}$ is symmetric we can instead apply the modified conjugate 
residual (MCR) solver adapted for indefinite systems~\cite{Ch77}, 
or the minimal residual (MINRES) method~\cite{PS75}, directly to 
the augmented system in \eqref{eq:lin_lin_aug}. If $\bm{L}$ 
is non-symmetric we can apply the bi-conjugate gradient 
stabilized (Bi-CGSTAB) solver~\cite{Vo92}, or the 
conjugate gradient for the normal equations (CGNR)~\cite{Bj96a, Sa03}.
An illustration of the above formulation in action can be seen in \Cref{fig:trui_harmonic_feat_comp}.
%


\subsection{Linear-Constrained Nonlinear Inpainting}
\label{sec:linear_constrained_nonlinear_inpainting}

For a nonlinear inpainting operator $\N:\mathbb{R}^N\to\mathbb{R}^N$ 
with linear equality constraints we can apply the previous result 
directly by analogy to 
\eqref{eq:nonlinear_inpainting_formulation_compact}, resulting in the 
following inpainting formulation:
\begin{equation}
    \bm{P}\bm{u}+(\bm{I}-\bm{P})\N(\bm{u}) = \bm{P}\bm{f}.
\end{equation}
A standard approach to solve this problem is to linearize $\N$ 
around some point $\bm{u}$
\begin{equation}
\N(\bm{u}+\bm{v}) = \N(\bm{u})+\bm{J}_{\N}(\bm{u})\bm{v} + 
O(\|\bm{v}\|^2),
\end{equation}
where $\bm{J}_{\N}$ is the Jacobian of $\N$. 
This results in the following Newton iteration:
\begin{equation}
\begin{aligned}
    \bm{P}\bm{u}^{k+1} + (\bm{I}-\bm{P})\N(\bm{u}^{k+1}) 
    &= \bm{P}\bm{f} \\
    \bm{P}(\bm{u}^k+\bm{v}^k)+(\bm{I}-\bm{P})
    (\N(\bm{u}^k)+\bm{J}_{\N}(\bm{u}^k)\bm{v}^k) &= \bm{P}\bm{f} \\
    (\bm{P}+(\bm{I}-\bm{P})\bm{J}_{\N}(\bm{u}^k))\bm{v}^k &=
    \bm{P}\bm{f} - \bm{P}\bm{u}^k - (\bm{I}-\bm{P})\N(\bm{u}^k).
\end{aligned}
\end{equation}
Its augmented system counterpart is
\begin{equation}
\label{eq:nlin_lin_aug}
    \begin{bmatrix}
        \bm{J}_{\N}(\bm{u}^k) & \bm{A}^{\top} \\
        \bm{A} & \bm{0}
    \end{bmatrix}
    \begin{bmatrix}
        \bm{v}^k \\ \bm{\lambda}^{k+1}
    \end{bmatrix}
    = \begin{bmatrix}
        -\N(\bm{u}^k) \\ \bm{A}\bm{f}-\bm{A}\bm{u}^k
    \end{bmatrix}.
\end{equation}
In both of the above formulations an issue becomes apparent if 
the projection of $\bm{J}_{\N}(\bm{u}^k)$ on the kernel of 
$\bm{A}$ is singular. \revhl{The system becomes inconsistent if 
$(\bm{I}-\bm{P})\N(\bm{u}^k)$ is not in the range of 
$(\bm{I}-\bm{P})\bm{J}_{\N}(\bm{u}^k$).} In practice, in such a case 
we can project $\N(\bm{u}^k)$ on 
$\spanV(\bm{J}_{\N}(\bm{u}^k))$, e.g.\ with CGNR, and 
then modify the right-hand side in \eqref{eq:nlin_lin_aug} 
to account for this.
Alternatively, one automatically gets the pseudoinverse solution 
if the augmented system \eqref{eq:nlin_lin_aug} is symmetric and 
tackled with the conjugate residual (CR) solver~\cite{Sa03} or 
MINRES solver~\cite{PS75} with the modification discussed 
in~\cite{LLR24}. If the system is non-symmetric then applying 
the conjugate gradient solver for 
the normal equations (CGNR)~\cite{Bj96a, Sa03}, with an initial 
guess with no components in the kernel (e.g.\ a zero initial guess), 
leads to the pseudoinverse solution~\cite{Ha20}.

We emphasize that in the above formulation $\N$ did not 
have to arise as the gradient of some energy, since we directly 
linearized the inpainting equation 
$\bm{P}\bm{u}+(\bm{I}-\bm{P})\N(\bm{u}) = \bm{P}\bm{f}$. This 
makes the method applicable to inpainting operators such as EED which 
do not have an energy~\cite{We21}. 

Under this framework, computing the Jacobian $\bm{J}_{\N}$  of the 
inpainting operator is necessary. While this is feasible using automatic 
differentiation, it can be computationally expensive, and may result 
in a non-symmetric matrix \revhl{if there exists no energy functional 
$E$ such that $\N= \bm{\nabla} E$}. 
In \Cref{sec:linear_constrained_quasilinear_inpainting} 
we show that 
a certain simplification is possible when the inpainting operator is 
quasi-linear (like EED), i.e., has the form 
$\N(\bm{u}) = \bm{L}(\bm{u})\bm{u}$. 
This allows avoiding the computation of the Jacobian, and if $\bm{L}(\bm{u})$ 
is symmetric it leads to a symmetric system.


\subsection{Linear-Constrained Quasi-Linear Inpainting}
\label{sec:linear_constrained_quasilinear_inpainting}

\begin{figure}[H]
    \begin{tabular}{ccc}
    \includegraphics[width=0.3\textwidth]{resources/trui/trui_c1_d0_05_EED3_ni30_ns1.jpg} &
    \includegraphics[width=0.3\textwidth]{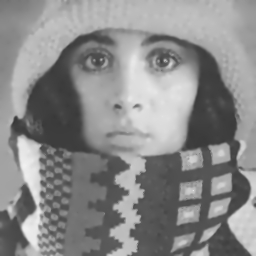} &
    \includegraphics[width=0.3\textwidth]{resources/trui/trui.jpg} \\
   gray values, MSE: $14.95$ & five features, MSE: $10.77$ & original image 
   \emph{trui} $\bm{f}$
    \end{tabular}
    \caption{Illustration of EED inpainting using only gray-value features versus five features (gray values, $\partial_x$, $\partial_y$, $3\times 3$ and $5\times 5$ binomial kernels). The total mask density is $5\%$ in both cases. \textbf{The reconstruction error improves by nearly $28\%$ compared to the already strong performance of classical EED inpainting}.}
    \label{fig:trui_EED_feat_comp}
\end{figure}

If the inpainting operator has the form 
$\N(\bm{u}) = \bm{L}(\bm{u})\bm{u}$, then we can avoid computing the 
Jacobian, by considering a Ka{\v c}anov iteration~\cite{Ka59} instead 
of a Newton iteration. 
The Ka{\v c}anov iteration is tantamount to dropping the term involving 
$\bm{J}_{\bm{L}}$ from the Jacobian $\bm{J}_{\N}$:
\begin{equation}
    \bm{J}_{\N}(\bm{u}) = \bm{L}(\bm{u}) + \bm{J}_{\bm{L}}(\bm{u})\bm{u} 
    \approx \bm{L}(\bm{u}).
\end{equation}
In practice this is applicable if $\bm{J}_{\bm{L}}$ does not change 
too quickly with $\bm{u}$, e.g., for EED inpainting this corresponds  
to a setting where the contrast parameter $\lambda$ is not too small.
Then the inpainting formulation has the simpler form
\begin{equation}
    (\bm{P}+(\bm{I}-\bm{P})\bm{L}(\bm{u}^k))\bm{u}^{k+1} = \bm{P}\bm{f},
\end{equation}
and its augmented system counterpart is
\begin{equation}
    \label{eq:qlin_lin_aug}
    \begin{bmatrix}
        \bm{L}(\bm{u}^k) & \bm{A}^{\top} \\
        \bm{A} & \bm{0}
    \end{bmatrix}
    \begin{bmatrix}
        \bm{u}^{k+1} \\ \bm{\lambda}^{k+1}
    \end{bmatrix}
    = \begin{bmatrix}
        \bm{0} \\ \bm{A}\bm{f}
    \end{bmatrix}.
\end{equation}
This has the additional benefit that if $\bm{L}(\bm{u}^k)$ is symmetric 
(as is the case for EED) then our problem remains symmetric, while there 
is no guarantee that the true Jacobian 
$\bm{J}_{\N}(\bm{u}) = \bm{L}(\bm{u}) + \bm{J}_{\bm{L}}(\bm{u})\bm{u}$ 
is symmetric (due to the second term). 
Moreover, we do not run into the potential issue where the 
system becomes potentially inconsistent if $\bm{J}_{\N}$ projected onto 
the kernel of $\bm{A}$ is singular, since we have a zero on the right-hand 
side. This means that solvers for consistent systems can be used, for example  
projected conjugate gradients or SYMMLQ~\cite{PS75} in the symmetric setting, 
and Bi-CGSTAB~\cite{Vo92} in the non-symmetric setting. An illustration 
of the results due to this formulation is shown in 
\Cref{fig:trui_EED_feat_comp}.
%


\section{Nonlinear Feature Inpainting}
\label{eq:nonlinear_features_inpainting}

Seeing as we were able to extend the feature interpolation framework 
to nonlinear inpainting operators, the natural question arises whether we 
can extend the framework to also handle nonlinear features. As demonstrated 
in this section, this is indeed possible.


\subsection{Optimization Formulation}

Let $\bm{b}:\mathbb{R}^N\to\mathbb{R}^M$ be a $C^2$ nonlinear 
function representing nonlinear constraints, and let the inpainting 
operator $\N(\bm{u})$ arise as 
the gradient $\N(\bm{u}) = \bm{\nabla} E(\bm{u})$ of some $C^2$ functional 
$E:\mathbb{R}^N\to\mathbb{R}$. We can formulate the inpainting  
as an equality constraint optimization problem
\begin{equation}
    \min_{\bm{b}(\bm{u})=0} E(\bm{u}) \implies 
    \min_{\bm{u}\in\mathbb{R}^N}\max_{\bm{\lambda}\in\mathbb{R}^M}
    \mathcal{L}(\bm{u},\bm{\lambda}), 
    \quad \mathcal{L}(\bm{u},\bm{\lambda}):= E(\bm{u}) 
    + \bm{\lambda}^{\top}\bm{b}(\bm{u}).
\end{equation}
A stationary point has to satisfy the first-order optimality conditions:
\begin{align}
    \label{eq:nabla_u_L}
    \bm{\nabla}_{\bm{u}}\mathcal{L}(\bm{u},\bm{\lambda}) &= 
    \bm{\nabla} E(\bm{u}) 
    + \bm{J}^{\top}_{\bm{b}}(\bm{u})\bm{\lambda} = \bm{0},
    \\
    \label{eq:nabla_lambda_L}
   \bm{\nabla}_{\bm{\lambda}} \mathcal{L}(\bm{u},\bm{\lambda}) 
   &= \bm{b}(\bm{u}) = \bm{0}.
\end{align}
By linearizing around $\bm{u}^k$ we get the augmented system
\begin{equation}
    \label{eq:SQP_augmented}
    \begin{bmatrix}
        \bm{\nabla}_{\bm{u}\bm{u}}\mathcal{L}(\bm{u}^k,\bm{\lambda}^k) & 
        \bm{J}_{\bm{b}}^{\top}(\bm{u}^k) \\
        \bm{J}_{\bm{b}}(\bm{u}^k) & \bm{0}
    \end{bmatrix}
    \begin{bmatrix}
        \bm{v}^k \\
        \bm{\lambda}^{k+1}
    \end{bmatrix}
    =
    -\begin{bmatrix}
        \bm{\nabla}E(\bm{u}^k) \\
        \bm{b}(\bm{u}^k)
    \end{bmatrix}.
\end{equation}
In the above we have that
\begin{equation}
    \bm{\nabla}E(\bm{u}) = \N(\bm{u}), \quad
    \bm{\nabla}_{\bm{u}\bm{u}}\mathcal{L}(\bm{u},\bm{\lambda}) =
    \bm{J}_{\N}(\bm{u}) + \bm{H}[\bm{b}](\bm{u})\bm{\lambda} =
     \bm{J}_{\N}(\bm{u}) +\sum_{j=1}^M \bm{H}[b_j](\bm{u}) \lambda_j,
\end{equation}
where $\bm{H}[b_j](\bm{u})\in\mathbb{R}^{N\times N}$ is the Hessian matrix of 
the $j$-th constraint function 
$b_j$ from the constraints vector $\bm{b}$. \Cref{eq:SQP_augmented} is the 
standard subproblem solved in each step of sequential quadratic 
programming (SQP)~\cite{NW06}. 


\subsection{Formulation in Terms of Nonlinear Equations}
\label{sec:nonlinear_constrained_nonlinear_inpainting}

We can be more general and not 
assume that $\N = \bm{\nabla} E$. Then we may start from 
\eqref{eq:nabla_u_L} and \eqref{eq:nabla_lambda_L} in the following form:
\begin{align} 
    \label{eq:nlin_inpainting_nlc_semi_compacta1}
    \N(\bm{u}) 
    + \bm{J}^{\top}_{\bm{b}}(\bm{u})\bm{\lambda} &= \bm{0},
    \\
    \label{eq:nlin_inpainting_nlc_semi_compacta2}
   \bm{b}(\bm{u}) &= \bm{0}.
\end{align}
The above is equivalent to the system
\begin{align} 
    \label{eq:nlin_inpainting_nlc_semi_compact1}
    (\bm{I}-\bm{P}(\bm{u}))\N(\bm{u}) &= \bm{0},
    \\
    \label{eq:nlin_inpainting_nlc_semi_compact2}
   \bm{b}(\bm{u}) &= \bm{0},
\end{align}
where $\bm{I}-\bm{P}(\bm{u})$ is the orthogonal projector on 
the kernel of the Jacobian $\ker(\bm{J}_{\bm{b}}(\bm{u}))$ of the constraints. 
Compare the above equations to \eqref{eq:lin_inpainting_semi_compact1} 
and \eqref{eq:lin_inpainting_semi_compact2}, as well as to
\eqref{eq:lin_inpainting_lc_semi_compact1} 
and \eqref{eq:lin_inpainting_lc_semi_compact2}. Setting 
$\bm{b}(\bm{u}) = \bm{A}\bm{u}-\bm{A}\bm{f}$ and 
$\N(\bm{u}) = \bm{L}\bm{u}$ recovers the linear setting.

The equations \eqref{eq:nlin_inpainting_nlc_semi_compact1} 
and \eqref{eq:nlin_inpainting_nlc_semi_compact2} make explicit the 
general form of our nonlinear inpainting formulation with nonlinear equality 
constraints. We enforce the constraints and require that $\N(\bm{u})$ 
is zero in the kernel of the Jacobian of the constraints. The linear operator 
projection formulation is recovered by linearizing the system, resulting 
in the projected Jacobian $(\bm{I}-\bm{P})\bm{J}_{\N}(\bm{I}-\bm{P})$. 
For large and sparse problems, such as ours, the more convenient 
formulation is the one using the augmented system
\begin{equation}
    \begin{bmatrix}
        \bm{J}_{\N}(\bm{u}^k)+\bm{H}[\bm{b}](\bm{u}^k)\bm{\lambda}^k & 
        \bm{J}_{\bm{b}}^{\top}(\bm{u}^k) \\
        \bm{J}_{\bm{b}}(\bm{u}^k) & \bm{0}
    \end{bmatrix}
    \begin{bmatrix}
        \bm{v}^k \\
        \bm{\lambda}^{k+1}
    \end{bmatrix}
    =
    -\begin{bmatrix}
        \N(\bm{u}^k) \\
        \bm{b}(\bm{u}^k)
    \end{bmatrix}.
\end{equation}
In the above it can happen that both parts of the right-hand side, 
$-\N(\bm{u}^k)$ and $-\bm{b}(\bm{u}^k)$, are not in the 
range of the system matrix. While one can use a solver such as CGNR 
to find a solution in the least squares sense, in practice it is 
more appropriate to project $\N(\bm{u}^k)$ on 
$\spanV(\bm{J}_{\N}(\bm{u}^k)+\bm{H}[\bm{b}](\bm{u}^k)\bm{\lambda}^k)$ and 
$\bm{b}(\bm{u}^k)$ on $\spanV(\bm{J}_{\bm{b}}(\bm{u}^k))$ and 
modify the right-hand side to contain these projections. 
Note that in the quasi-linear setting, 
$\N(\bm{u}) = \bm{L}(\bm{u})\bm{u}$, and if the approach from 
\Cref{sec:linear_constrained_quasilinear_inpainting} is employed,
 then inconsistency of the system cannot arise with respect to 
$\N(\bm{u}^k) = \bm{L}(\bm{u}^k)\bm{u}^k$ unless 
$\bm{H}[\bm{b}](\bm{u}^k)\bm{\lambda}^k$ reduces the rank of $\bm{L}(\bm{u}^k)$. 
If $\bm{L}$ is positive semi-definite and $\bm{H}[\bm{b}](\bm{u}^k)\bm{\lambda}^k$ 
is positive semi-definite this cannot occur because 
$\rank(\bm{A}+\bm{B})\geq \max(\rank(\bm{A}),\rank(\bm{B}))$ for positive 
semi-definite matrices.


\subsection{Solution Strategies}
\label{sec:solution_strategies}

In the setting of linear constraints with a quasi-linear 
inpainting operator, as in 
\Cref{sec:linear_constrained_quasilinear_inpainting}, solving the 
inpainting problem is fairly straightforward as one needs only 
a few Ka{\v c}anov steps. Solving 
the resulting linear systems~\eqref{eq:qlin_lin_aug} can be achieved by 
using, e.g., the conjugate residual solver~\cite{Sa03}. Nonlinear constraints make 
the problem much more challenging, as the normal and tangent spaces change 
from iteration to iteration.

\paragraph{Line Search vs Trust Region}
Newton iterations are not globally convergent in general. 
As a remedy one typically considers globalization through line 
search or trust region methods~\cite{CGT00,NW06}. In our problem we have found
that, without additional regularization, line search SQP 
approaches fail to converge even when considering theoretically well motivated 
acceptance criteria~\cite{BCN08}. This seems 
to be partially due to $\bm{J}_{\bm{b}}$ becoming rank deficient, and partially 
because the Newton directions for ill-conditioned problems become close to 
orthogonal to the gradient~\cite{TWS02}. 
Forcing sequences resulting in loose tolerances in the solvers~\cite{EW96} 
can somewhat mitigate this, but the SQP method may still fail to 
converge.

\paragraph{Byrd-Omojokun Step Decomposition}

Trust region approaches with Steihaug-Krylov~\cite{St83} solvers are more 
robust for our problem but can also stall. The quasi-normal 
step from the Byrd-Omojokun's trust region approach 
(for an overview see \cite{LNP98} and \cite{NW06}) is crucial 
in our setting for the generation of a reliable direction. 
The main idea is to solve a minimization problem for the constraints 
\begin{equation}
    \min_{\bm{v}^k_N\in\mathcal{C}^k}\|\bm{v}_N^k\|^2_2, \,\,\textrm{ such that } \,\, \|\bm{v}^k_N\|_2 \leq \xi \Delta_k; 
    \quad \mathcal{C}^k = \argmin_{v} \|\bm{b}(\bm{u}^k)+\bm{J}_{\bm{b}}(\bm{u}^k)\bm{v}\|^2_2, 
\end{equation}
resulting in the quasi-normal direction $\bm{v}^k_N$. The full step is 
$\bm{v}^k_N = -\bm{J}_{\bm{b}}^+(\bm{u}^k)\bm{b}(\bm{u}^k)$, however, the 
trust region constraint $\|\bm{v}^k_N\|_2 \leq \xi \Delta_k$ may require 
a modification of the latter. Having made a step towards feasibility, one 
can then focus on a direction $\bm{v}_T^k$  purely targeted at improving the 
objective in the tangent space of the constraints
\begin{equation}
    \label{eq:SQP_tangential_subpreoblem}
    \begin{bmatrix}
        \bm{J}_{\N}(\bm{u}^k)+\bm{H}[\bm{b}](\bm{u}^k)\bm{\lambda}^k & 
        \bm{J}_{\bm{b}}^{\top}(\bm{u}^k) \\
        \bm{J}_{\bm{b}}(\bm{u}^k) & \bm{0}
    \end{bmatrix}
    \begin{bmatrix}
        \bm{v}^k_N+\bm{v}^k_T \\
        \bm{\lambda}^{k+1}
    \end{bmatrix}
    =
    \begin{bmatrix}
        -\N(\bm{u}^k) \\
         \bm{J}_{\bm{b}}(\bm{u}^k)\bm{v}^k_N
    \end{bmatrix}, \quad
    \|\bm{v}_N^k+\bm{v}_T^k\|_2 \leq \Delta_k.
\end{equation}
The system is now consistent, at least with respect to the normality
conditions.

\paragraph{Model and Merit Functions}
In practice we implement both trust region constraints 
by using our Steihaug-CGNR solver, which terminates when the step violates 
the trust region radius. Since CGNR monotonically decreases the 
residual norm, a corresponding model function is
\begin{equation}
    m_k(\bm{v}) = 
    \min_{\bm{\lambda}}\|(\bm{J}_{\N}(\bm{u}^k)+\bm{H}[\bm{b}](\bm{u}^k))\bm{v} 
    + \bm{J}_{\bm{b}}^{\top}(\bm{u}^k)\bm{\lambda}\|^2_2 
    + \rho^2 \|\bm{b}(\bm{u}^k) + \bm{J}_{\bm{b}}(\bm{u}^k)\bm{v}\|^2_2.
\end{equation}
The merit function is based on the residuals of the 
original nonlinear equations \eqref{eq:nlin_inpainting_nlc_semi_compacta1} 
and \eqref{eq:nlin_inpainting_nlc_semi_compacta2}:
\begin{equation}
    \varphi_k(\bm{v}) = \min_{\bm{\lambda}}\|\N(\bm{u}^k+\bm{v}) 
    + \bm{J}^{\top}_{\bm{b}}(\bm{u}^k+\bm{v})\bm{\lambda}\|^2_2 
    + \rho^2 \|\bm{b}(\bm{u}^k+\bm{v})\|^2_2.
\end{equation}
In practice we omit the minimization over $\bm{\lambda}$ and replace 
$\bm{\lambda}$ with $\bm{\lambda}^{k+1}$ in $m_k(\bm{v}^k)$ and 
$\varphi_k(\bm{v}^k)$, and with $\bm{\lambda}^k$ in 
$m_k(\bm{0})=\varphi_k(\bm{0})$. In our experiments $\bm{\lambda}^k$ 
becomes closer and closer to the optimal $\bm{\lambda}$ as 
the method converges.

\paragraph{Trust Region Update}
Our approach to step acceptance or rejection, as well as trust region 
expansion or contraction, follows classical methods~\cite{CGT00,NW06}, and 
is based on the ratio of actual reduction $ared$ to predicted reduction $pred$:
\begin{equation}
    r_k = \frac{ared}{pred} = 
    \frac{\varphi_k(\bm{0})-\varphi_k(\bm{v}^k)}{m_k(\bm{0})-m_k(\bm{v}^k)}
    \implies
    \Delta_{k+1} \in 
    \begin{cases}
        \gamma_2 \Delta_k, &\text{ for }
        r_k\geq \eta_2 \land \|\bm{v}^k\|=\Delta_k, \\
        \Delta_k, &\text{ for }r_k\in [\eta_1,\eta_2), \\
        \gamma_1 \Delta_k, &\text{ for }r_k<\eta_1.
    \end{cases}
\end{equation}
In the above $0<\eta_1\leq \eta_2 \leq 1$ are thresholds to 
determine whether a step is to be accepted or rejected, and whether the 
trust region is to be grown or shrunk:
\begin{itemize}
    \item $r_k<\eta_1$: the step is rejected and the trust radius is shrunk by $\gamma_1$,
    \item $r_k\in[\eta_1,\eta_2)$: the step is accepted and the trust region  remains the same,
    \item $r_k\geq \eta_2$: the step is accepted; the trust radius is grown by $\gamma_2$ only if it was interfering with the step, i.e., if $\|\bm{v}^k\|=\Delta_k$.
\end{itemize}

\section{Spatial Optimization}
\label{sec:spatial_optimization}

In the current section we discuss our spatial optimization approach 
for selecting the data from which the image is to be reconstructed. 
The latter is crucial for a good quality reconstruction.
We extend the algorithm from our previous work~\cite{JCW23} 
to be able to handle nonlinear features and inpaintings. Additionally, there 
are two modifications that boost its performance 
considerably even in the linear setting 
(see the last paragraph of \Cref{sec:exp_harmonic}). 
We also provide connections to matching pursuit~\cite{MZ93} and 
sensing dictionaries~\cite{SV08} and thus better motivate 
some of the choices made in the design of the algorithm.

\subsection{Problem Formulation in the Classical Setting}

Given a set of features and a data budget, the \emph{spatial optimization} 
has the goal of selecting a subset of features as interpolation constraints 
within this budget, such that the best possible reconstruction is achieved. 
In the simplest case of gray value interpolation this can be 
formulated as the following sparse approximation problem:
\begin{equation}
    \label{eq:spatial_optimization_classical}
    \min_{\|\bm{c}\|_0 \leq m} \|\bm{u}(\bm{c},\bm{C}\bm{f})-\bm{f}\|^2_2, 
    \quad \|\bm{c}\|_0 = \sum_{i=1}^N [c_i\ne 0],
\end{equation}
where $m$ is the maximum number of mask points provided for by our data 
budget, and $\bm{u}(\bm{c},\bm{C}\bm{f})$ is the reconstruction based 
on the mask $\bm{c}$ and the interpolation data $\bm{C}\bm{f}$.
Already in the simple case of only interpolating gray values, we have  
a combinatorial problem of size 
\begin{equation}
{N\choose m} = O\left(\left(\frac{e \cdot N}{m}\right)^m\right).
\end{equation}
As such it is intractable in practice except for very small $N$ 
(note that in practice $m$ is typically a fixed fraction of $N$). 
Therefore, efficient algorithms such as matching pursuit~\cite{MZ93} 
and its variants have been developed for finding a suboptimal but good 
enough solution for a simpler variant of this problem. We provide  
more details on the relationship to matching pursuit in 
\cref{seq:relation_to_synthesis}.

\subsection{Problem Formulation in the Setting of Linear Features}

As an example, we consider feature matrices 
that represent color value interpolation $\bm{F}_1 = \bm{I}$, 
$x$ derivatives $\bm{F}_2\approx \partial_x$, 
$y$ derivatives $\bm{F}_3\approx \partial_y$, 
and local integrals $\bm{F}_4 \approx K_{\sigma} * $. Then the 
selection process can be formulated in terms of a larger mask 
$\bm{c} = (\bm{c}_1,\bm{c}_2,\bm{c}_3,\bm{c}_4)\in\{0,1\}^{4N}$
made up of the four masks for the four features
\begin{equation}
    \min_{\|\bm{c}\|_0 \leq m} \|\bm{u}(\bm{c},\bm{A}\bm{f})-\bm{f}\|^2_2, 
    \,\, \bm{A}\bm{u} = \bm{A}\bm{f}; \,\,
    \bm{A} = \bm{C}\bm{F} = 
    \begin{bmatrix}
        \bm{C}_1 & & & \\
        &\bm{C}_2 && \\ &&\bm{C}_3& \\ &&&\bm{C}_4
    \end{bmatrix}
    \begin{bmatrix}
        \bm{F}_1 \\ \bm{F}_2 \\ \bm{F}_3 \\ \bm{F}_4
    \end{bmatrix}
    =
    \begin{bmatrix}
        \bm{C}_1\bm{F}_1 \\ \bm{C}_2\bm{F}_2 \\
        \bm{C}_3\bm{F}_3 \\ \bm{C}_4\bm{F}_4
    \end{bmatrix}.
\end{equation}
The four different masks describe the pixel locations in the image 
at which we wish to interpolate the four different features. \revhl{It is allowed 
for those to overlap in the image domain}, e.g., to interpolate both
color values and derivatives at the same pixel.
Note also that $\bm{A}\in\mathbb{R}^{M\times N}$ was allowed to be non-square 
in our constrained inpainting formulation. From the above 
we conclude that if we consider $q$ different types of features, the search 
space grows to ${q\cdot N \choose m}$, since we have to pick $m$ rows out 
of the $M=q\cdot N$ rows of the matrix $\bm{F}$. For the 
optimization approach that we consider, this growth of the search space is 
not a major issue.


\subsection{Problem Formulation in the Setting of Nonlinear Features}

We can write both linear and nonlinear features as a constraint 
$\bm{\phi}(\bm{u})-\bm{\phi}(\bm{f}) = \bm{0}$.  Suppose we have $q$ families 
of potentially nonlinear constraints $\bm{\phi}_1,\ldots,\bm{\phi}_q:\mathbb{R}^N\to\mathbb{R}^N$ defined 
for every pixel. Given 
$q$ masks $\bm{c}_1,\ldots,\bm{c}_q$, where $c_{ij}=1$ means 
that constraint $i$ is active at pixel $j$, we can formulate the 
constraint function $\bm{b}:\mathbb{R}^N\to\mathbb{R}^{q\cdot N}$ as follows:
\begin{equation}
    \bm{b}(\bm{u}) = \bm{C}\bm{\phi}(\bm{u})-\bm{C}\bm{\phi}(\bm{f}), \quad 
    \bm{C}\bm{\phi} = 
    \begin{bmatrix}
        \bm{C}_1 & &\\
        & \ddots & \\
        & & \bm{C}_q
    \end{bmatrix}
    \begin{bmatrix}
        \bm{\phi}_1 \\ \vdots \\ \bm{\phi}_q
    \end{bmatrix}
    =
    \begin{bmatrix}
        \bm{C}_1\bm{\phi}_1 \\ \vdots \\ \bm{C}_q\bm{\phi}_q
    \end{bmatrix}.
\end{equation}
Then the minimization problem can be formulated concisely as 
\begin{equation}
    \min_{\|\bm{c}\|_0 \leq m} \|\bm{u}(\bm{c},\bm{C}\bm{\phi}(\bm{f}))-\bm{f}\|^2_2,
    \,\, \textrm{ such that }\,\, \bm{b}(\bm{u}) = 
    \bm{C}\bm{\phi}(\bm{u})-\bm{C}\bm{\phi}(\bm{f}) = \bm{0}.
\end{equation}
Note that in the linear setting $\bm{\phi}_j(\bm{u}) = \bm{F}_j\cdot \bm{u}$, 
where $\cdot$ is matrix-vector multiplication.

\subsection{Densification and Sparsification}

An efficient approach to tackling the spatial optimization problem involves iteratively adding or removing mask points over multiple iterations. In densification approaches we start with an empty mask, 
and at each iteration we aim to add mask points with locations chosen 
such that the error is decreased the most. In sparsification approaches 
we start from a full mask, and aim to remove mask points in each iteration, 
such that the error increases the least. Generally densification approaches 
are a more natural choice if we aim at a mask density under $50\%$, while 
sparsification approaches are more efficient for mask densities over $50\%$. 
\revhl{There are, however, other considerations. For instance, adding points
usually makes the influence of previously selected points more local, 
whereas removing points makes their influence more global.} 
Taking this into account and also the fact that we target 
low densities, we choose to use a densification method.

\paragraph{Computational Cost of Full Densification (Sparsification)} 
In the extreme case we can add (remove) a single point per iteration 
and evaluate the error for all possible candidates. 
This means that in step $k$ we need to perform as many 
inpaintings as there are zeros (ones) in the mask $\bm{c}^{k}$ in order 
to evaluate the error. When summed over all steps, this yields a total of $m!$ 
(sparsification: $(N-m)!$) number of inpaintings. To reduce this, we 
add (remove) multiple mask points per iteration, and evaluate the inpainting 
only at some locations instead of all possible locations. Furthermore, 
instead of computing an inpainting for all of the candidates and recomputing 
the MSE, we instead use an oracle that tries to predict which are the best 
candidates. We discuss the oracle in \cref{sec:error_map} and a way 
to avoid clustering when adding multiple mask points in 
\cref{sec:partition_based_densification}.


\subsection{Error Map}
\label{sec:error_map}

In each densification iteration we compute the \textit{error map}  
$\bm{e}^k = \bm{u}^k-\bm{f}$ in order to try and predict at what locations 
new mask points should be introduced. If we normalize the rows 
of the matrix $\bm{J}_{\bm{\phi}}(\bm{f})$ 
w.r.t.\ the $2$-norm, and denote the latter as $\tilde{\bm{J}}_{\bm{\phi}}(\bm{f})$, 
the entries of the matrix-vector product 
$\tilde{\bm{J}}_{\bm{\phi}}(\bm{f})\cdot \bm{e}^k$ give 
us the dot products between the normalized linearizations of the features 
and the error. The normalization of the rows is done so that different 
feature types are weighted in a similar fashion. 
We can expect that at locations where the magnitude of these inner 
products is large, the specific feature can compensate well for the error. 
This is similar in nature to the sensing step in matching pursuit~\cite{MZ93} 
or thresholding~\cite{SV08}, however, in our case we do not have the 
atoms giving the true reconstruction.

In the linear setting $\bm{\phi}^k(\bm{e}^k) = \bm{J}_{\bm{\phi}}(\bm{f})\cdot \bm{e}^k$ 
so we have a one-to-one correspondence between the features and the matrix 
$\bm{J}_{\bm{\phi}}(\bm{f})$. In the nonlinear setting using 
$\bm{J}_{\bm{\phi}}(\bm{f})$ is motivated by the following expansion:
\begin{equation}
    \bm{\phi}(\bm{u}^k) = \bm{\phi}(\bm{f}+\bm{e}^k)
    \approx \bm{\phi}(\bm{f}) + \bm{J}_{\bm{\phi}}(\bm{f}) \cdot \bm{e}^k
    \implies 
     \bm{\phi}(\bm{u}^k) - \bm{\phi}(\bm{f}) \approx 
     \bm{J}_{\bm{\phi}}(\bm{f}) \cdot (\bm{u}^k-\bm{f}).
\end{equation}
As noted, unlike the classical matching pursuit setting, we do not use the 
true reconstruction atoms -- the inpainting echoes -- since the latter are 
prohibitively expensive to compute~\cite{GW25}.  
Consequently, the magnitudes of the entries of 
$\tilde{\bm{J}}_{\bm{\phi}}(\bm{f})\cdot \bm{e}^k$ 
typically underestimate the error decrease. Notably, they ignore the 
additional decrease in a local neighborhood due to the inpainting operator. 
Moreover, if we try to introduce multiple points per iteration, there 
is a danger that they cluster due to trying to reduce the same error. 
We mitigate both of these issues in the next subsection.

\subsection{Partition-Based Densification}
\label{sec:partition_based_densification}

\begin{figure}
    \centering
    \begin{tabular}{ccc}
        \includegraphics[width=0.3\linewidth]{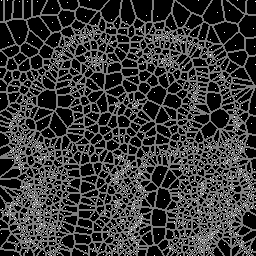}
        &
        \includegraphics[width=0.3\linewidth]{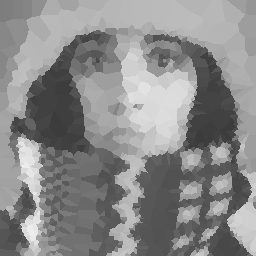}
        &
        \includegraphics[width=0.3\linewidth]{resources/trui/trui.jpg} \\
        Voronoi tessellation & piecewise constant approx.\ & 
        original image \emph{trui}
    \end{tabular}
    \caption{A visualization of the Voronoi tessellation induced by 
    a mask, a piecewise-constant approximation, 
    and the original image.}
    \label{fig:Voronoi_viz}
\end{figure}

In order to disallow clustering when introducing multiple points we 
partition the domain and allow at most one point to be introduced 
per region. A partition is non-overlapping by definition and 
it should ideally be adapted to the current mask structure. A simple 
partition with a multitude of desirable properties is given by the Voronoi 
tessellation~\cite{AKL13} -- a visualization of such a partition 
adapted to a mask is shown in \Cref{fig:Voronoi_viz}.
In order to mitigate the issue of $\tilde{\bm{J}}_{\bm{\phi}}(\bm{f})\cdot\bm{e}^k$ 
underestimating the error decrease, we additionally integrate the squared error 
in each cell and choose to introduce new mask points only in the 
$\lceil m/n \rceil$ cells with highest errors ($n$ being the number of densification 
iterations). 
This works under the assumption that locally the inpainting 
operator is able to decrease the error inside the cell. 
The location where we add the new mask point within the cell and its feature 
type are determined by the maximal pointwise error 
$\tilde{\bm{J}}_{\bm{\phi}}(\bm{f})\cdot\bm{e}^k$ within the cell.

\subsection{Voronoi Densification for \revhl{Equality Constraint Features}}

Putting all of these ideas together yields an improved and generalized 
variant of the algorithm from our conference publication~\cite{JCW23}. We employ 
a densification strategy with an inpainting at each iteration, 
an error map computed from the latter, and a Voronoi decomposition. 
The main difference is the support for nonlinear features. 
We also now normalize the rows of $\bm{J}_{\bm{\phi}}(\bm{f})$ 
w.r.t.\ the $2$-norm instead of the $1$-norm previously used.
Another change from~\cite{JCW23} is that we integrate 
the squared error per cell, instead of the squared features' errors. 
The listed modifications are more in line with well-founded ideas from  
matching pursuit algorithms and yield considerable 
improvements. \Cref{alg:voronoi-densification} briefly describes the main 
steps. The initial mask $\bm{c}^1$ in the algorithm can be random or 
can be the result of halftoning 
the image $|\N(\bm{f})|$. The latter is a generalization of the 
idea from Belhachmi et al.'s publication~\cite{BBBW09}.

\begin{algorithm}[htb]
	\caption{Voronoi Densification for Feature Inpainting} 
	\label{alg:voronoi-densification} 
    \textbf{Input     :} Original image $\bm{f}$,  
    number of iterations $n$, number of desired mask points $m$  \\
    \textbf{Output    :} Inpainting mask $\bm{c}^n$, reconstruction $\bm{u}$\\
    \textbf{Initialize:} Initial mask $\bm{c}^1$ with 
    $\lceil \frac{m}{n} \rceil$ mask pixels
    \begin{algorithmic}[1]
        \FOR{$k=1$ \TO $n-1$}
            \STATE
            Construct the Voronoi tessellation $\{\mathcal{T}_{j}\}$ 
            of the current mask pixels.
            \STATE
            Compute the inpainting 
            $\bm{u}^k = \bm{u}(\bm{c}^k,\bm{C}^k\bm{\phi}(\bm{f}))$
            and the error map $\bm{e}^k=\bm{u}^k-\bm{f}$. 
            \STATE
            Compute the cells $2$-norm errors
            $\forall j, \, e^k_{\mathcal{T}_j} = \sum_{i\in \mathcal{T}_j} (e^k_i)^2$.
            \STATE
            Find the $\lceil\frac{m}{n}\rceil$ Voronoi cells
            $\{\mathcal{T}_{j_i}\}_{i=1}^{\lceil \frac{m}{n} \rceil}$ with the highest 
            errors $\{e^k_{\mathcal{T}_{j_i}}\}_{i=1}^{\lceil\frac{m}{n}\rceil}$.
            \STATE
            For each cell in $\{\mathcal{T}_{j_i}\}_{i=1}^{\lceil\frac{m}{n}\rceil}$ 
            find the entry in $\tilde{\bm{J}}_{\bm{\phi}}(\bm{f})\cdot\bm{e}^k|_{\mathcal{T}_{j_i}}$ with
            highest magnitude and add a mask point to the corresponding 
            location in $\bm{c}^{k+1}$.
        \ENDFOR
 \end{algorithmic}
\end{algorithm}



\subsection{Relation to Matching Pursuit}
\label{seq:relation_to_synthesis}

In the literature, matching pursuit~\cite{MZ93,RS09} algorithms have been used 
to solve a simpler and somewhat different version of the combinatorial 
problem in \eqref{eq:spatial_optimization_classical}, namely
\begin{equation}
    \min_{\bm{a};\|\bm{c}\|_0\leq m}\|\bm{D}\bm{C}\bm{a}-\bm{f}\|^2_2,
\end{equation}
where $\bm{D}$ is a dictionary, $\bm{a}$ are coefficients, and $\bm{c}$
is the mask. 
The main difference compared to our problem is that the dictionary is given, 
and that the reconstruction $\bm{u} = \bm{D}\bm{C}\bm{a}$ is linear in 
the coefficients. The above can in fact be reconciled with linear 
inpainting with color value interpolation constraints by considering 
the spectral counterpart of the inpainting problem~\cite{HPW15}. 
That is, if we have an 
inpainting operator $\bm{L}$, we may consider its pseudoinverse 
\begin{equation}
    \bm{L}^+ = 
    \begin{bmatrix}
        \bm{W}_{\img} & \bm{W}_{\ker}
    \end{bmatrix}
    \begin{bmatrix}
        \bm{\Sigma}_{11}^{-1} & \bm{0} \\
        \bm{0} & \bm{0}
    \end{bmatrix}
    \begin{bmatrix}
        \bm{U}_{\img}^{\top} \\ \bm{U}_{\ker}^{\top}
    \end{bmatrix}, \quad
    \bm{L} = 
    \begin{bmatrix}
        \bm{U}_{\img} & \bm{U}_{\ker}
    \end{bmatrix}
    \begin{bmatrix}
        \bm{\Sigma}_{11} & \bm{0} \\
        \bm{0} & \bm{0}
    \end{bmatrix}
    \begin{bmatrix}
        \bm{W}_{\img}^{\top} \\ \bm{W}_{\ker}^{\top}
    \end{bmatrix},
\end{equation}
then any inpainting can be written as
\begin{equation}
    \bm{u} = \bm{L}^+\bm{C}\bm{a} + \bm{W}_{\ker}\bm{\mu}, \quad 
    \bm{U}_{\ker}^{\top}\bm{C}\bm{a} = \bm{0}.
\end{equation}
Here $\bm{L}^+$ plays the role of the dictionary and 
$\bm{a}$ plays the role of the coefficients. We also need 
a part from the kernel, but the optimal values for the latter 
can be computed directly as $\bm{\mu} = \bm{W}^{\top}_{\ker}\bm{f}$.  
Then we can write the problem in a similar way as our classical 
formulation
\begin{equation}
    \underset{\substack{\|\bm{c}\|_0\leq m\\
    \bm{U}_{\ker}^{\top}\bm{C}\bm{a}=\bm{0}}}{\min}\|\bm{L}^+\bm{C}\bm{a}-
    \bm{W}_{\img}\bm{W}_{\img}^{\top}\bm{f}\|^2_2.
\end{equation}
Applying matching pursuit can work well provided $\bm{L}^+$ is 
not too ill-conditioned. If that is the case (which happens to 
be so even for harmonic inpainting), one typically needs to consider 
orthogonal matching pursuit which also performs a projection at each step. 
This has been done in 1D for harmonic 
and biharmonic inpainting in the work by Plonka et al.~\cite{PHW16}. 
Note that harmonic inpainting in 1D is linear interpolation, while 
biharmonic inpainting in 1D is cubic interpolation. 

In either case, it is clear that in the 2D setting performing 
a projection at each step can prove to be prohibitively 
expensive. Moreover, it is not directly clear how the linear feature 
formulation can be brought to a form such as (or if this is feasible at all) 
\begin{equation}
    \bm{u} = \bm{D}\bm{C}\bm{a} + \bm{W}_{\ker}\bm{\mu}.
\end{equation}
An idea would be to consider $\bm{D} = (\bm{F}\bm{L})^+$ (where 
$\bm{M}^+$ is the Moore-Penrose inverse), which 
results in a synthesis and not an analysis approach, but 
we do not pursue this further here, as it also does not generalize to 
nonlinear inpainting operators and nonlinear features.

\section{Experiments}
\label{sec:experiments}	


After discussing the experimental setup in \cref{sec:exp_setup}, 
we provide visual and qualitative experiments in \cref{sec:exp_harmonic} 
for harmonic inpainting, and in \cref{sec:exp_EED} for EED inpainting. 
We briefly mention the setting of nonlinear features in \cref{exp:nlin_nlin}, 
demonstrating that the framework is also valid in the nonlinear case.


\subsection{Experimental Setup}
\label{sec:exp_setup}

\paragraph{Image Test Set}
We conduct experiments on twelve $512\times 512$ natural images that exhibit 
a wide range of frequencies and structural patterns; 
see \cref{fig:exp_collected_harmonic} and \cref{fig:exp_collected_EED}.
For each image, we perform $30$ densification iterations, using different 
sampling densities tailored to each case. The densities are chosen to ensure 
that the error remains noticeable, allowing for a meaningful visual comparison 
between inpaintings with different feature sets. 

\revhl{The only exceptions are the density scaling 
experiment and the experiment for the nonlinear features.
For the scaling experiment we use \textit{trui} with $n=30$.}
For the nonlinear feature experiments we also use $30$ 
densification iterations on a $64\times 64$ checkerboard. Natural images 
do not contain a sufficiently large percentage of corners for the effect 
to be obvious, and thus we have chosen the checkerboard image.


\paragraph{Linear Features}
\revhl{We define five linear feature types $\bm{\phi}_1,\ldots,\bm{\phi}_5$ 
through the convolutions:
\begin{equation}
    (\bm{\phi}_1(\bm{v}))_{i,j} = v_{i,j}, \quad 
    (\bm{\phi}_2(\bm{v}))_{i,j} = v_{i+1,j}-v_{i,j}, 
    \quad 
     (\bm{\phi}_2(\bm{v}))_{i,j} = v_{i,j+1}-v_{i,j}, 
\end{equation}
\begin{equation}
    \bm{\phi}_4(\bm{v})= \begin{bmatrix}
        0 & 0 & 0 & 0 & 0 \\
        0 & 1 & 2 & 1 & 0 \\
        0 & 2 & 4 & 2 & 0 \\
        0 & 1 & 2 & 1 & 0 \\
        0 & 0 & 0 & 0 & 0 \\
    \end{bmatrix}*\bm{v},  \quad 
    \bm{\phi}_5(\bm{v}) = \begin{bmatrix}
        1 &  4 &  6 &  4 & 1 \\
        4 & 16 & 24 & 16 & 4 \\
        6 & 24 & 36 & 24 & 6 \\
        4 & 16 & 24 & 16 & 4 \\
        1 &  4 &  6 &  4 & 1 \\
    \end{bmatrix}*\bm{v}.
\end{equation}}
The first feature is the identity and serves to interpolate 
color values like in the classical setting, the second and third describe
discrete derivatives, and the fourth and fifth represent 
discrete counterparts of Gaussian convolution. We note that the stencils 
are automatically normalized by our spatial optimization so multiplicative 
constants have no effect on the densification.

In our experiments, we evaluate the impact of progressively increasing the 
number of considered feature families -- see \Cref{tbl:MSE_table_harmonic} 
and \Cref{tbl:MSE_table_EED}. In these tables, the index $q$ in the leftmost 
column denotes the inclusion of the first $q$ feature families, i.e., the 
set $\{\bm{\phi}_k\}_{k=1}^q$. 
For example, $q=3$ corresponds to using color values along with
interpolation of the $x$- and $y$-derivatives. \textbf{We note that 
throughout all of our experiments, whenever we compare 
reconstructions with different feature types, we use the same total 
mask density, except where explicitly stated otherwise.}


\paragraph{Nonlinear Feature}
In the nonlinear setting we replace $\bm{\phi}_2$ with an affine-invariant 
curvature feature $\tilde{\phi_2}$ (see \cite{SNOW98} and the references therein):
\begin{equation}
    \label{eq:weighted_variance_feature}
    (\tilde{\bm{\phi}}_2)_{i,j}(\bm{u}) \approx (u_y^2 u_{xx} - 2u_x u_y u_{xy} + u_x^2 u_{yy})(x_i,y_j),
\end{equation}
where we discretise all derivatives with central differences:
\begin{align}
    (u_x)(x_i,y_j) &= \frac{u_{i+1,j}-u_{i-1,j}}{2h_x} + O(h_x^2), \\
    (u_y)(x_i,y_j) &= \frac{u_{i,j+1}-u_{i,j-1}}{2h_y} + O(h_y^2), \\
    (u_{xx})(x_i,y_j) &= \frac{u_{i+1,j}-2u_{i,j}+u_{i-1,j}}{h^2_x} + O(h_x^2), \\
    (u_{yy})(x_i,y_j) &= \frac{u_{i,j+1}-2u_{i,j}+u_{i,j-1}}{h^2_y} + O(h_y^2), \\
    (u_{xy})(x_i,y_j) &= \frac{u_{i+1,j+1}-u_{i-1,j+1}-u_{i+1,j-1}+u_{i-1,j-1}}{4h_xh_y} + O(h_x^2+h_y^2).
\end{align}
%
\paragraph{Harmonic Inpainting}

For harmonic inpainting we use the standard $5$-point stencil 
discretization of the negated Laplacian:
\begin{equation}
    (-\Delta u)(x_i,y_j) \approx (\bm{L}\bm{u})_{ij} = 
    \frac{-u_{i-1,j}+2u_{i,j}-u_{i+1,j}}{h_x^2} + 
    \frac{-u_{i,j-1}+2u_{i,j}-u_{i,j+1}}{h_y^2},
\end{equation}
with reflecting boundary conditions: $u_{0,j} := u_{1,j}$, 
$u_{W+1,j} := u_{W,j}$, $u_{i,0}:=u_{i,1}$, $u_{i,H+1} := u_{i,H}$.

\paragraph{EED Inpainting}
The matrix for EED inpainting is constructed according to the 
discretization in~\cite{WWW13}. We use the following parameters: 
contrast parameter $\lambda=1.0$, standard deviation for 
Gaussian pre-smoothing $\sigma=0.8$, parameters $\alpha=0.25$ and  
$\gamma=1.0$ from~\cite{WWW13}. For images that contain higher 
frequencies such as \textit{shed, raindeer, madeira,} and \textit{crab}, 
we use a lower pre-smoothing parameter $\sigma=0.1$, which results in  
better quality.
At each step of the densification process, our EED inpainting 
implementation initializes with the harmonic inpainting solution 
and subsequently applies Ka{\v c}anov iterations to compute the 
final result.
%


\paragraph{Inpaintings with Linear Constraints}

For the linear systems arising in linearly constrained harmonic and EED 
inpainting we apply the modified conjugate residual solver~\cite{Ch77} to 
\eqref{eq:lin_lin_aug} and \eqref{eq:qlin_lin_aug}.
We use a stopping criterion based on the residual 
$\|\bm{r}\|_2\leq 10^{-8}\|\tilde{\bm{r}_0}\|_2$ where 
\begin{equation}
    \|\tilde{\bm{r}}_0\|_2 = \|(\bm{I}-\bm{P})\bm{L}(\bm{u}^k)\bm{P}\bm{f}\|_2
\end{equation}
is the residual norm corresponding to an initial guess that satisfies the 
constraints. 
The matrix–vector products with the projection $\bm{P}=\bm{A}^+\bm{A}$ are 
computed using the conjugate gradient method for the normal equations 
(CGNR)~\cite{Sa03}. The iteration is initialized with the zero vector 
and terminated once the relative residual reaches $10^{-12}$. This procedure 
yields an approximation of the Moore–Penrose pseudoinverse 
solution~\cite{Ha20}.


\paragraph{Inpaintings with Nonlinear Constraints}

In the setting of nonlinear equality constraints we use the SQP 
approach discussed in \cref{sec:solution_strategies}. As an initial 
guess to the SQP iteration we take the solution from harmonic inpainting 
with only the linear equality constraints being active. In each SQP step 
we use our Steihaug-CGNR solver for the computation of the quasi-normal and 
tangential directions. 
We have found the latter to be the most robust for SQP, e.g., compared to 
other solvers such as the modified conjugate residual solver. The stopping 
criteria are based on the relative residual of the normal equations 
$\|\bm{M}\bm{r}\|_2\leq 10^{-8}\|\bm{M}\bm{r}_0\|_2$, where $\bm{M}$ is 
the system matrix in \eqref{eq:SQP_tangential_subpreoblem} for 
the tangential step, and for the quasi-normal step 
we have $\bm{M} = \bm{J}_{\bm{b}}^{\top}(\bm{u}^k)$.

\subsection{Harmonic Inpainting with Linear Features}
\label{sec:exp_harmonic}

We present a series of experiments demonstrating the robustness of our 
densification approach and its significant improvement over previous 
results in sparse harmonic inpainting.

\begin{figure}
    \begin{tabular}{ccc}
    \includegraphics[width=0.3\textwidth]{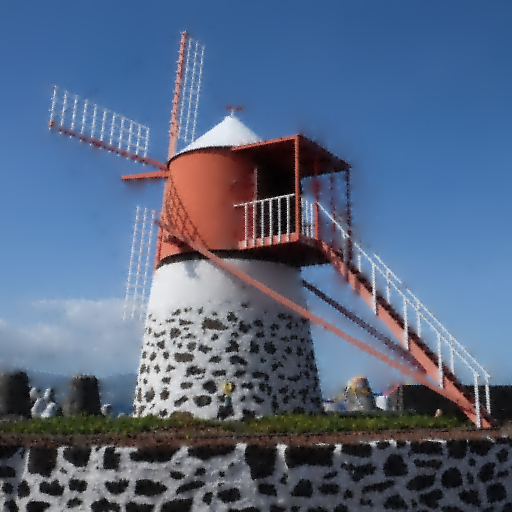} &
    \includegraphics[width=0.3\textwidth]{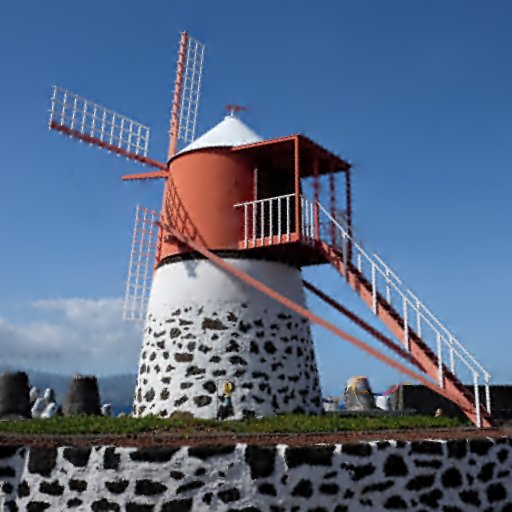} &
    \includegraphics[width=0.3\textwidth]{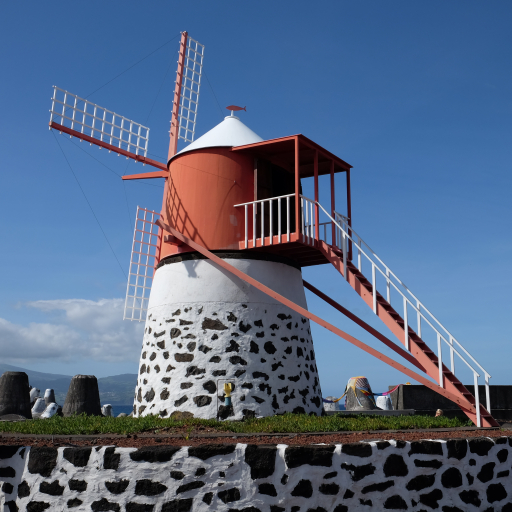} \\
    color values only & all five features & original
    \end{tabular}
    \caption{Harmonic inpainting with linear features, 5\% mask. Note the 
    large improvements in the closing of the edges and the suppression of the logarithmic singularities inherent to harmonic inpainting.}
    \label{fig:exp_visual_widmill}
\end{figure}

\begin{figure}
\begin{center}
    \begin{tabular}{ccc}
    \includegraphics[width=0.3\textwidth]{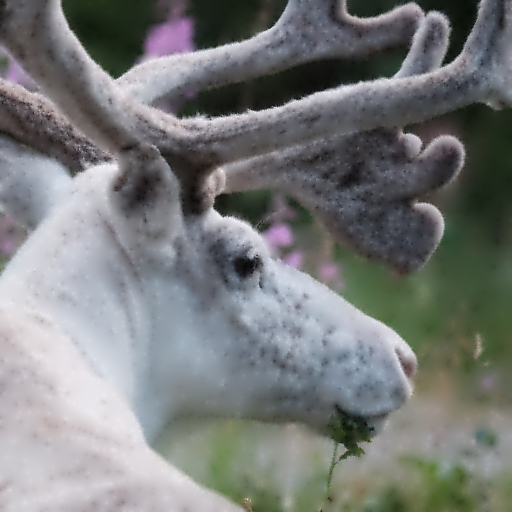} 
    \hspace{-2.55mm} & \hspace{-2.55mm}
    \includegraphics[width=0.3\textwidth]{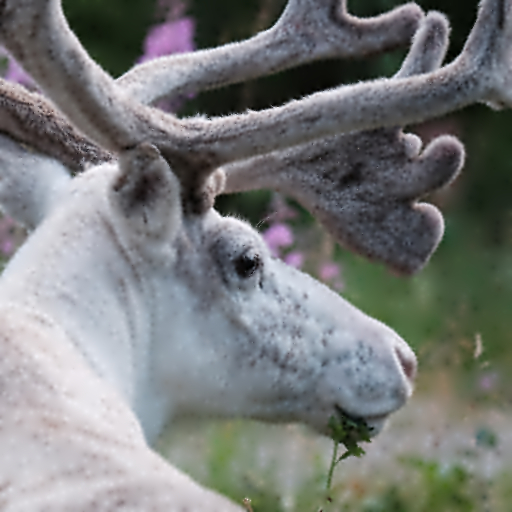}
    \hspace{-2.55mm} & \hspace{-2.55mm}
    \includegraphics[width=0.3\textwidth]{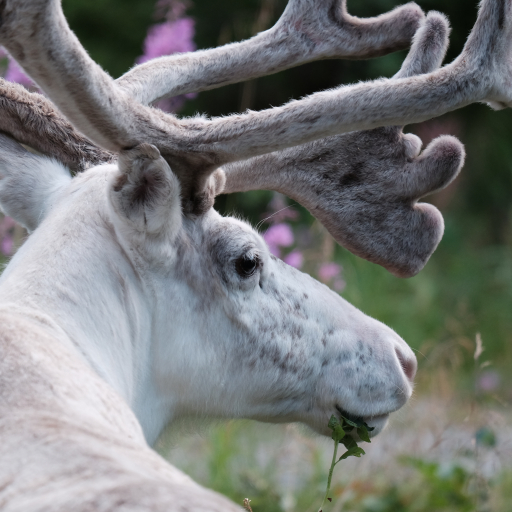}  
    \\
    \includegraphics[width=0.3\textwidth]{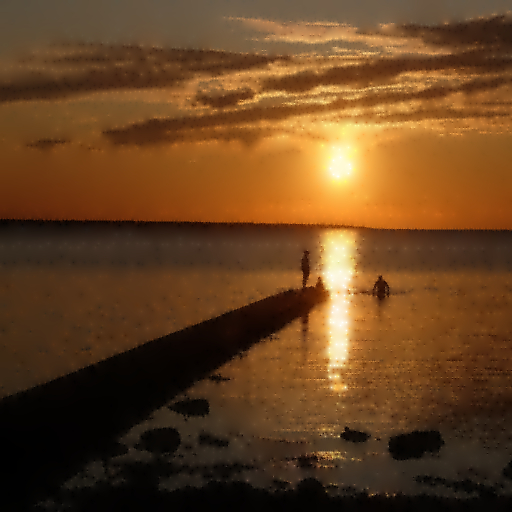}
    \hspace{-2.55mm} & \hspace{-2.55mm}
    \includegraphics[width=0.3\textwidth]{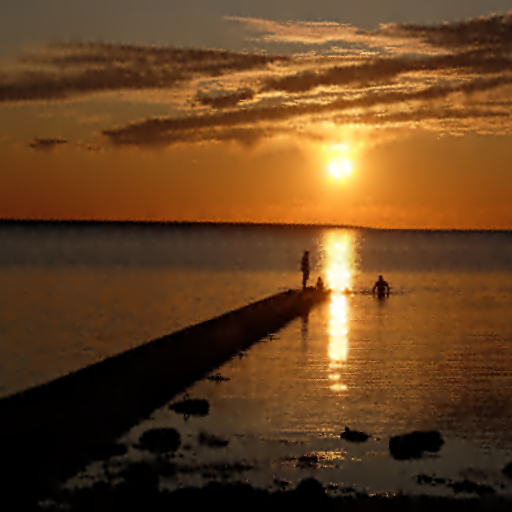}
    \hspace{-2.55mm} & \hspace{-2.55mm}
    \includegraphics[width=0.3\textwidth]{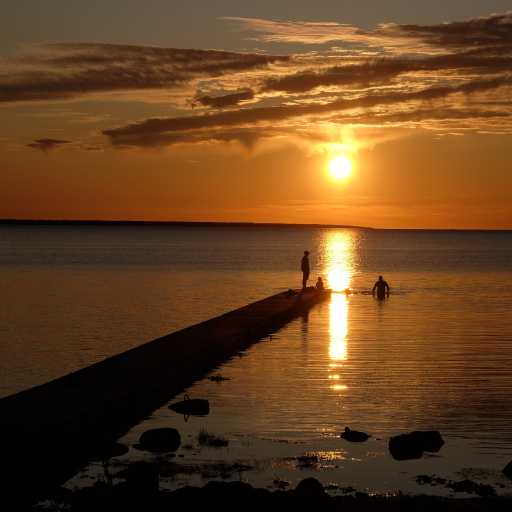}  
    \\
    \includegraphics[width=0.3\textwidth]{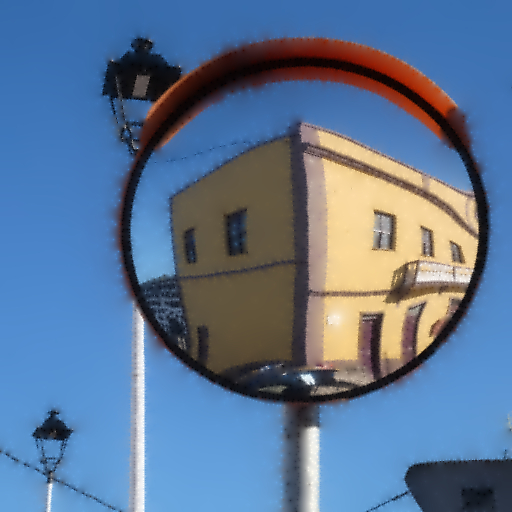} 
    \hspace{-2.55mm} & \hspace{-2.55mm}
    \includegraphics[width=0.3\textwidth]{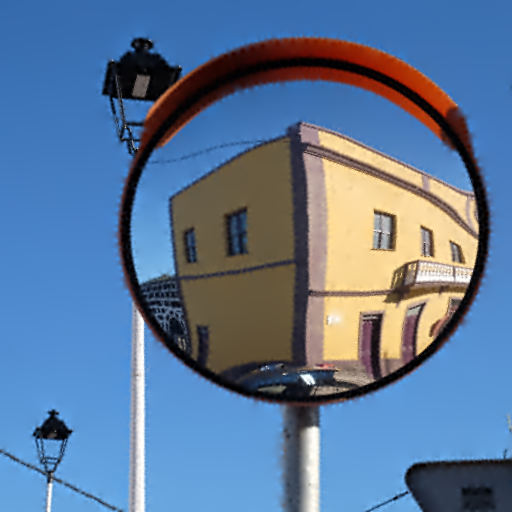}
    \hspace{-2.55mm} & \hspace{-2.55mm}
    \includegraphics[width=0.3\textwidth]{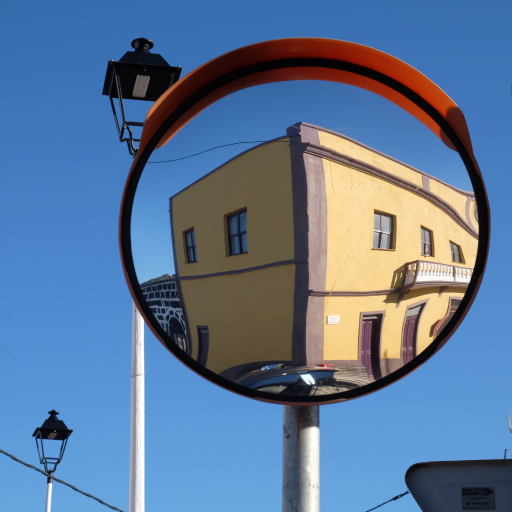}  
    \\
    \includegraphics[width=0.3\textwidth]{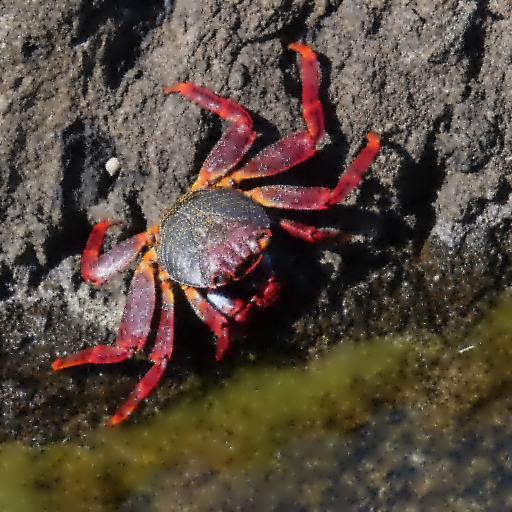} 
    \hspace{-2.55mm} & \hspace{-2.55mm}
    \includegraphics[width=0.3\textwidth]{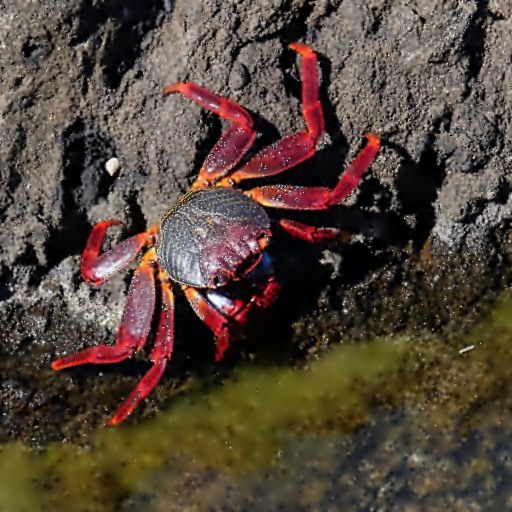} 
    \hspace{-2.55mm} & \hspace{-2.55mm}
    \includegraphics[width=0.3\textwidth]{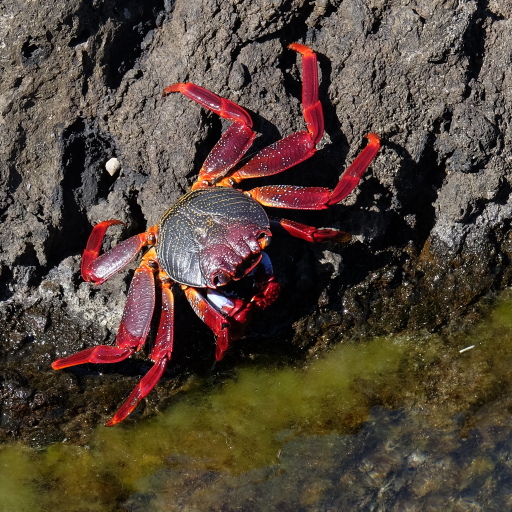}  
    \\
    \end{tabular}
\end{center}
    \caption{Harmonic inpainting. Left: color values only; 
    middle: all five features; right: original. Photos 
    by Joachim Weickert -- top to bottom: 
    \textit{raindeer, quai, mirror, crab}.}
    \label{fig:exp_collected_harmonic}
\end{figure}

\paragraph{Visual Comparison}
\Cref{fig:exp_visual_widmill} showcases the large difference between 
harmonic inpainting only with color interpolation and with the five 
linear features. In both cases we use the exact same mask 
density. We note that the classical harmonic inpainting is 
both much blurrier and suffers from pronounced logarithmic singularities. 
\revhl{These singularities are inherited from the continuous setting, where 
the Green's function of the Laplacian in two dimensions is 
$G(r)=\frac{1}{2\pi}\ln r$, with $r$ denoting the distance to the 
source point. Visually, these 
singularities appear as sharp local extrema around the specified mask pixels, 
see the leftmost image in \Cref{fig:exp_visual_widmill}.}
More visual comparisons can be found in 
\cref{fig:exp_collected_harmonic}. We highly recommend zooming 
into the images to see the difference. They are 
most obvious when one switches between the images on a computer.

\paragraph{Mean Squared Error}
\Cref{tbl:MSE_table_harmonic} lists the MSEs for all twelve images for 
an increasing set of feature types $\{\bm{\phi}_k\}_{k=1}^{q}$ 
while using the exact same total mask density. Note that the original 
images have range $\{0,1,\ldots,255\}$. We are able to decrease 
the MSE to be one third for the image \textit{mirror}, and often we are 
able to halve the MSEs. We get a very significant \textbf{average peak 
signal-to-noise ratio (PSNR) improvement of $\bm{2.76}$ dB} when comparing 
the setting with all five features ($q=5$) and only one feature ($q=1$). 

We note that the most beneficial feature is the $3\times 3$ binomial 
kernel $\bm{\phi}_4$. 
As expected, the $5\times 5$ binomial kernel $\bm{\phi}_5$ has little 
effect once $\bm{\phi}_4$ is present, since it is just a variant of 
discrete Gaussian convolution with a larger standard 
deviation than for $\bm{\phi}_4$. 
The derivative features $\bm{\phi}_2,\,\bm{\phi}_3$ have a more subtle 
effect than the $3\times 3$ binomial feature $\bm{\phi}_4$ 
on natural images -- they are much better suited to cartoon-like or 
piecewise-constant images as discussed in our conference paper~\cite{JCW23}.

\revhl{
\paragraph{Density Scaling in Terms of MSE}
For the image \textit{trui}, we additionally consider the inverse
relation between mask density and reconstruction quality.
\Cref{tbl:trui_harmonic_scaling} reports the densities required to achieve
selected target MSE values with harmonic inpainting. Doubling the 
target MSE reduces the required mask density to roughly $60\%$ 
of its previous value, while using five feature families instead of a 
single feature requires only about $44\%$--$56\%$ of the density 
needed to attain the same reconstruction quality. The relative
 savings increase as the mask density decreases.}

%
%
\begin{table}
\begin{center}
\begin{tabular}{|l|c|c|c|c|c|c|}
\hline
$q$ & \textit{boats} 7\% & \textit{elpaso} 3\% & \textit{shed} 15\% & 
\textit{generator} 3\% & \textit{raindeer} 3\% & \textit{quai} 3\% \\
\hline
1 & 144.71 & 76.03 & 110.27 & 175.66 & 57.21 & 67.22 \\
\hline
3 & 126.22 & 62.88 &  91.82 & 164.29 & 55.81 & 62.47 \\
\hline
4 &  66.53 & 34.76 &  80.20 & 100.52 & 35.47 & 41.46\\
\hline
5 &  \textbf{65.26} & \textbf{33.85} &  \textbf{79.41} &  \textbf{99.48} & \textbf{34.55} & \textbf{39.59} \\
\hline
\hline 
$q$ & \textit{mirror 3\%} & \textit{madeira 8\%} & \textit{garafia 7\%} & 
\textit{flowers 3\%} & \textit{crab 8\%} & \textit{windmill 5\%} \\
\hline
1 & 127.39 & 75.09 & 134.43 & 195.40 & 332.53 & 184.91 \\
\hline 
3 & 120.32 & 66.88 & 122.16 & 192.56 & 292.70 & 161.65 \\
\hline 
4 &  44.98 & 59.08 &  70.00 &  87.46 & 240.72 &  81.20 \\
\hline 
5 &  \textbf{43.01} & \textbf{58.53} &  \textbf{68.16} &  \textbf{79.17} & \textbf{238.12} &  \textbf{78.94} \\
\hline
\end{tabular}
\end{center}
\caption{MSEs for harmonic inpainting  with an increasing number of 
feature families.}
\label{tbl:MSE_table_harmonic}
\end{table}
%
%

\begin{table}
\centering
\begin{tabular}{|l|c|c|c|c|}
\hline
target MSE & 20 & 40 & 80 & 160 \\
\hline
1 feature
& 7.55\%
& 4.88\%
& 3.07\%
& 1.84\% \\
\hline
5 features
& 4.23\% 
& 2.42\%
& 1.35\%
& 0.82\%  \\
\hline
\end{tabular}
\caption{\revhl{Mask density in percent required to reach the specified 
target MSE for a harmonic inpainting of \textit{trui}.
 Doubling the target MSE reduces the required 
mask density by approximately a factor of $0.6$. 
Using five features reduces the required density by 
$44\%$--$56\%$.}}
\label{tbl:trui_harmonic_scaling}
\end{table}
%

\paragraph{Inpaintings and Masks}

\begin{figure}
\center
    {
    \setlength{\tabcolsep}{1pt}
    \begin{tabularx}{\textwidth}{>{\columncolor{gray!10}}c >{\columncolor{gray!10}}c 
    >{\columncolor{gray!10}}c >{\columncolor{gray!10}}c >{\columncolor{gray!10}}c}
    	\multirow{1}{*}[13ex]{\rotatebox[origin=c]{90}{inpainting}} &
    	\includegraphics[width=0.22\textwidth]
    		{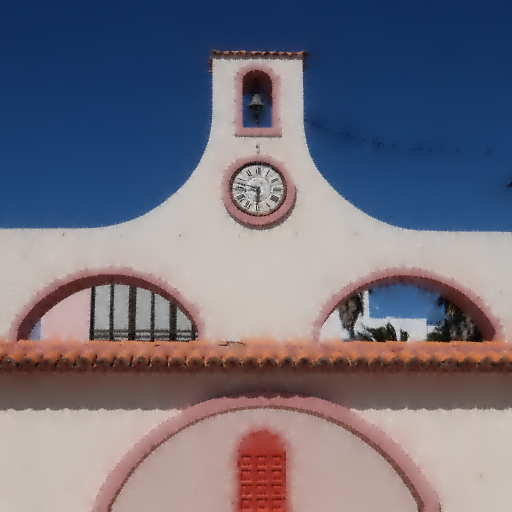} &
    	\includegraphics[width=0.22\textwidth]
    		{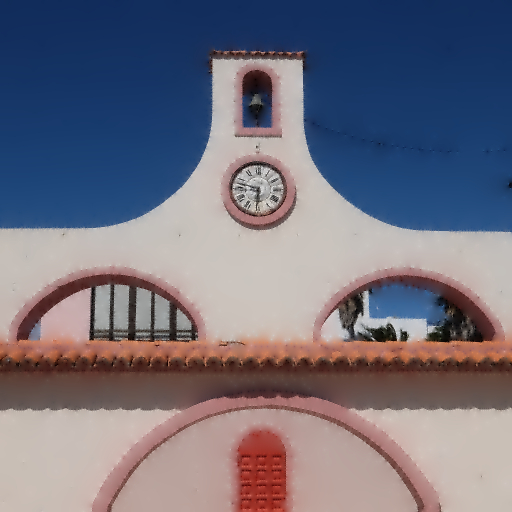} &
    	\includegraphics[width=0.22\textwidth]
    		{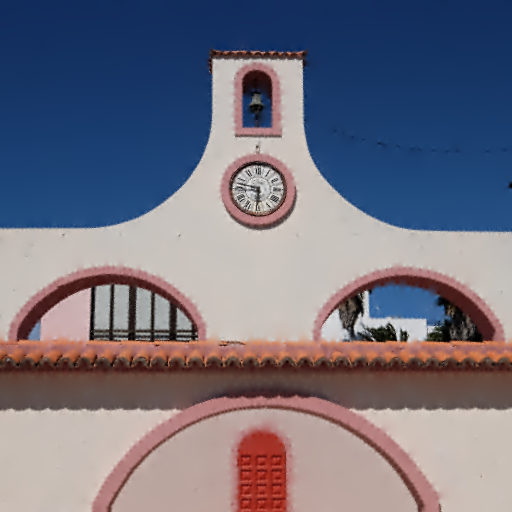} &
    	\includegraphics[width=0.22\textwidth]
    		{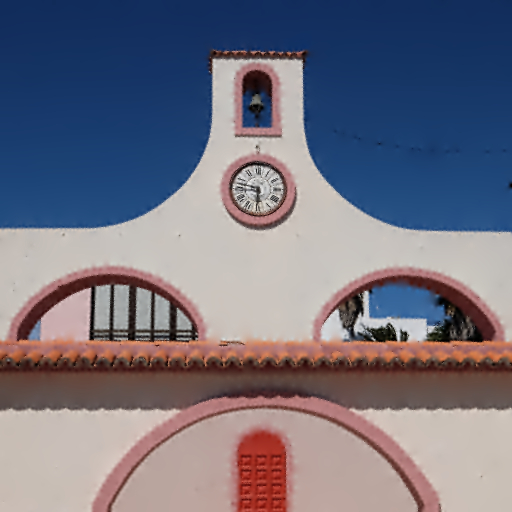} \\
    	
    	\multirow{1}{*}[8ex]{\rotatebox[origin=c]{90}{color}} &
    	\includegraphics[width=0.22\textwidth]
    		{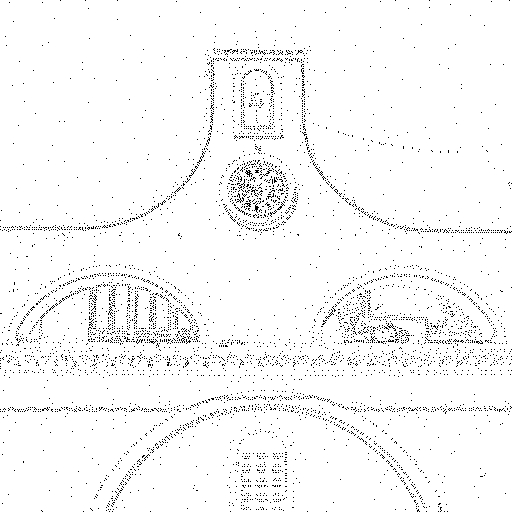} &
    	\includegraphics[width=0.22\textwidth]
    		{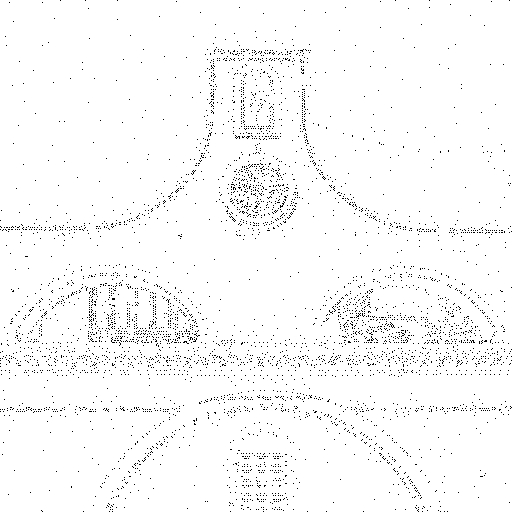} &
    	\includegraphics[width=0.22\textwidth]
    		{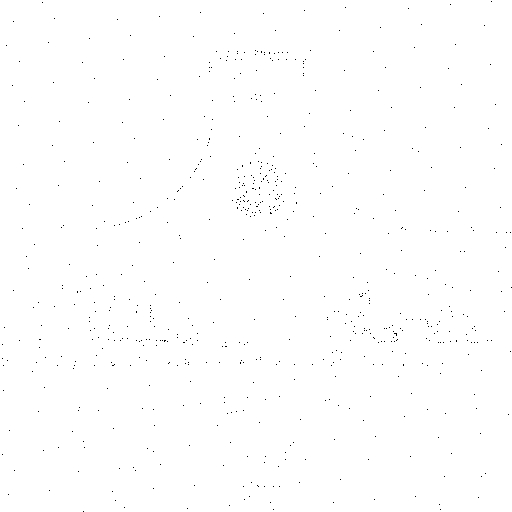} &
    	\includegraphics[width=0.22\textwidth]
    		{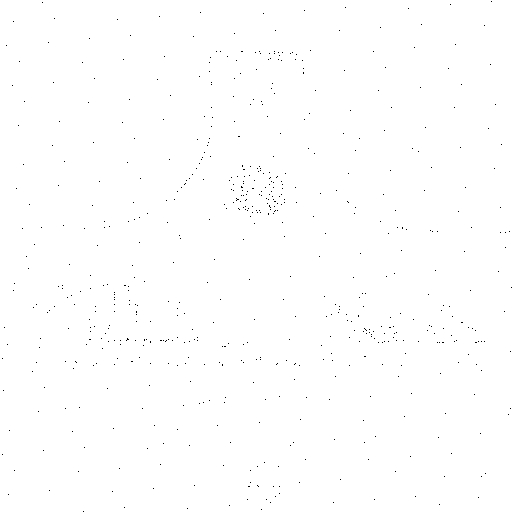} \\
    	
    	\multirow{1}{*}[7ex]{\rotatebox[origin=c]{90}{$\partial_x$}} &
    	~ &
    	\includegraphics[width=0.22\textwidth]
    		{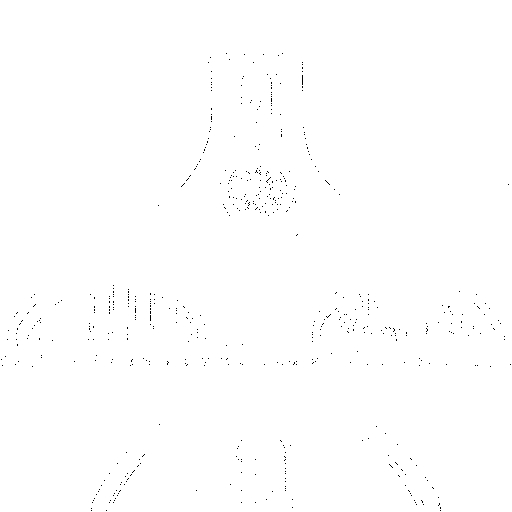} &
    	\includegraphics[width=0.22\textwidth]
    		{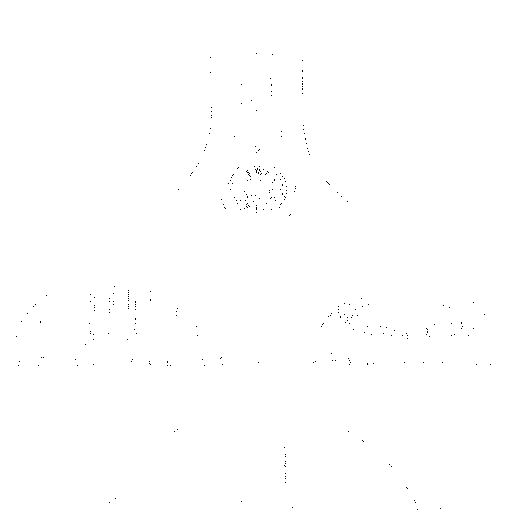} &
    	\includegraphics[width=0.22\textwidth]
    		{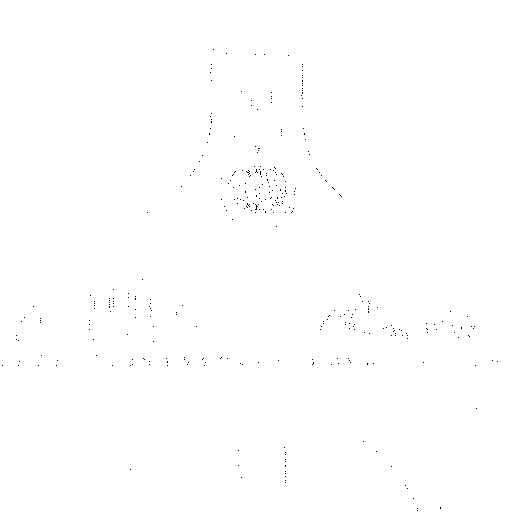} \\
    	
    	\multirow{1}{*}[7ex]{\rotatebox[origin=c]{90}{$\partial_y$}} &
    	~ &
    	\includegraphics[width=0.22\textwidth]
    		{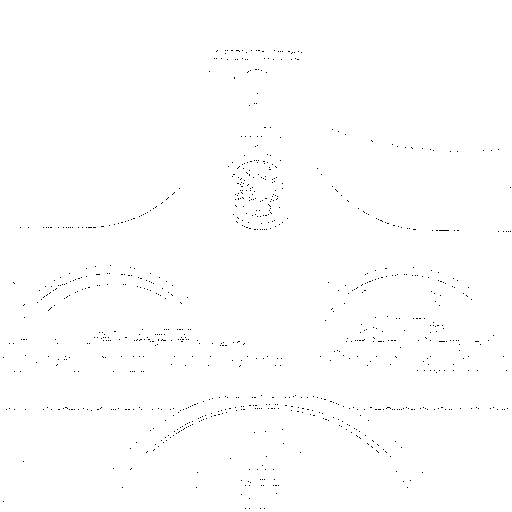} &
    	\includegraphics[width=0.22\textwidth]
    		{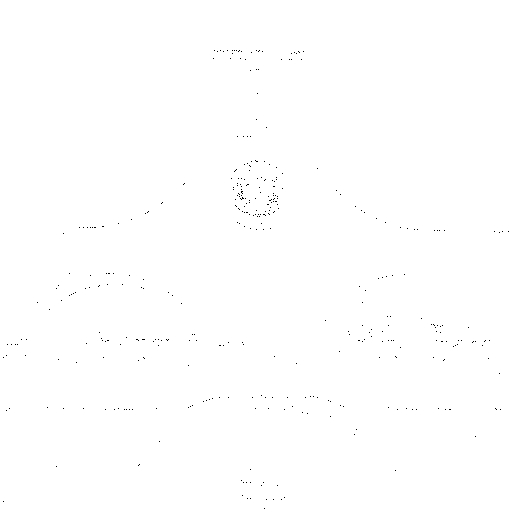} &
    	\includegraphics[width=0.22\textwidth]
    		{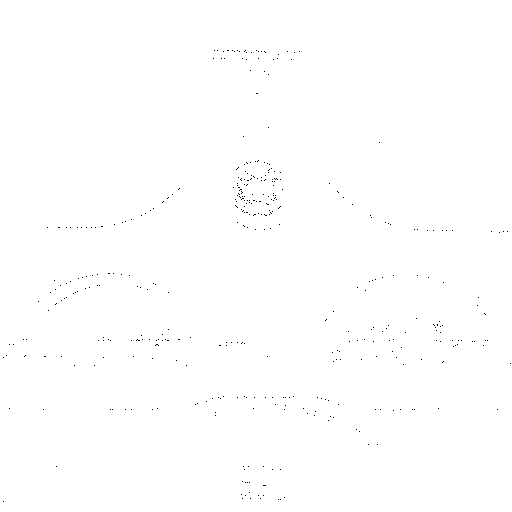}  \\
    	
    	\multirow{1}{*}[9ex]{\rotatebox[origin=c]{90}{$3 \times 3$}} &
    	~ &
    	~ &
    	\includegraphics[width=0.22\textwidth]
    		{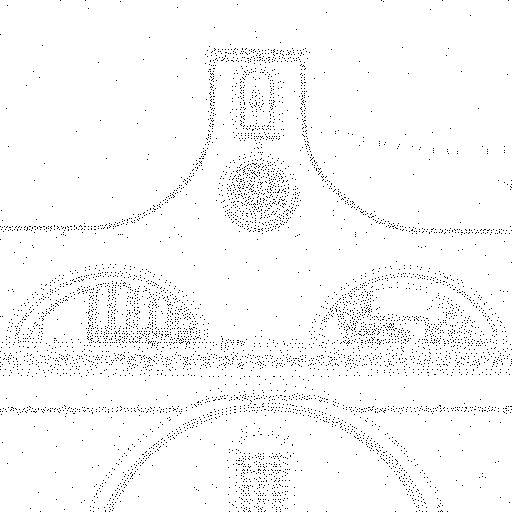} &
    	\includegraphics[width=0.22\textwidth]
    		{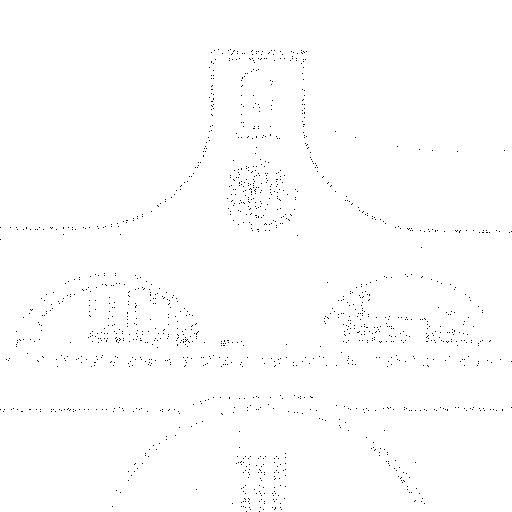} \\
    	
    	\multirow{1}{*}[9ex]{\rotatebox[origin=c]{90}{$5 \times 5$}} &
    	~ &
    	~ &
    	~ &
    	\includegraphics[width=0.22\textwidth]
    		{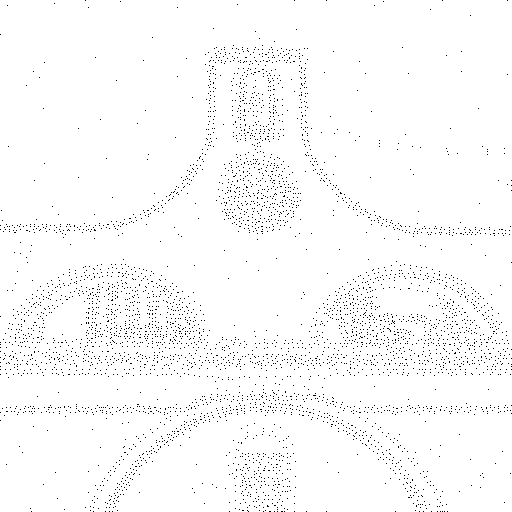} \\
    	
    	~ &
    	MSE:~$76.03$ &
    	MSE:~$62.88$ &
    	MSE:~$34.76$ &
    	MSE:~$\bm{33.85}$ \\
    	
    \end{tabularx}
    }
    \caption{Harmonic inpainting results for a combined density of $3\%$ 
    for \emph{elpaso} produced with 30 densification iterations.}
    \label{fig:exp_qualitative_elpaso}
\end{figure}

\begin{figure}
\center
    {
    \setlength{\tabcolsep}{1pt}
    \begin{tabularx}{\textwidth}{>{\columncolor{gray!10}}c >{\columncolor{gray!10}}c 
    >{\columncolor{gray!10}}c >{\columncolor{gray!10}}c >{\columncolor{gray!10}}c}
    	\multirow{1}{*}[13ex]{\rotatebox[origin=c]{90}{inpainting}} &
    	\includegraphics[width=0.22\textwidth]
    		{resources/harmonic/windmill/windmill-512_c3_d0_05_EED0_ni30_ns1.jpg} &
    	\includegraphics[width=0.22\textwidth]
    		{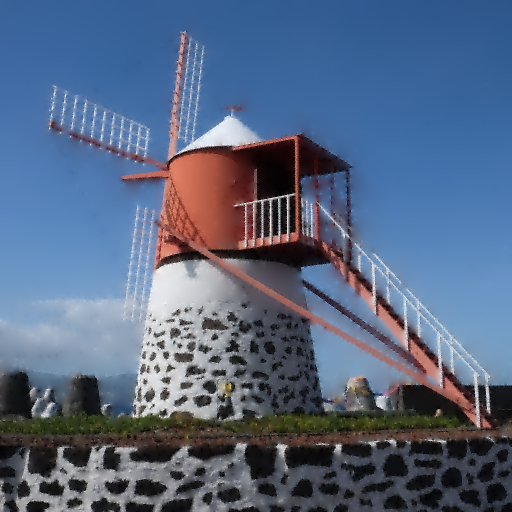} &
    	\includegraphics[width=0.22\textwidth]
    		{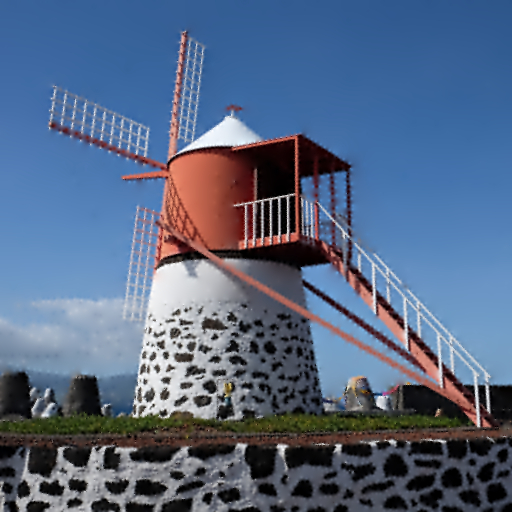} &
    	\includegraphics[width=0.22\textwidth]
    		{resources/harmonic/windmill/windmill-512_c3_d0_05_EED0_ni30_ns5.jpg} \\
    	
    	\multirow{1}{*}[8ex]{\rotatebox[origin=c]{90}{color}} &
    	\includegraphics[width=0.22\textwidth]
    		{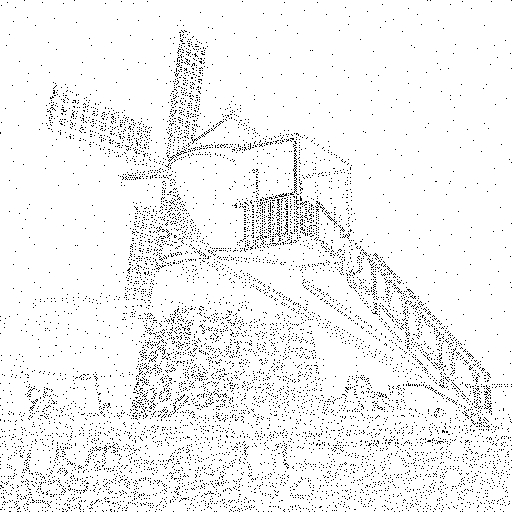} &
    	\includegraphics[width=0.22\textwidth]
    		{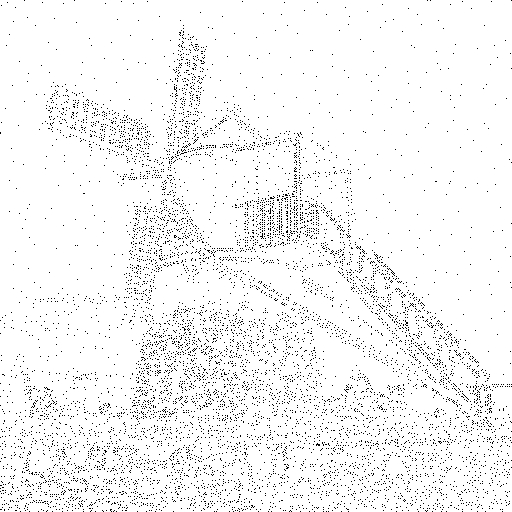} &
    	\includegraphics[width=0.22\textwidth]
    		{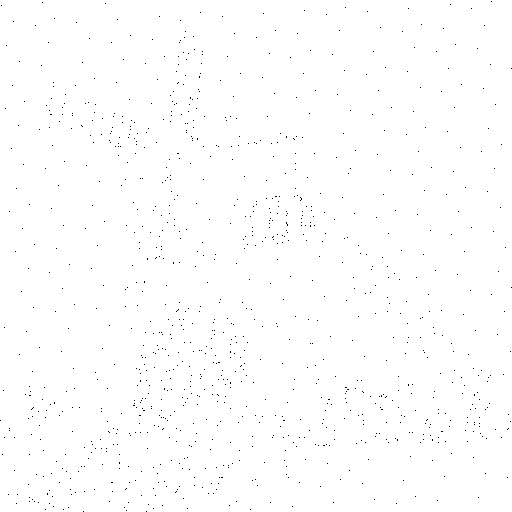} &
    	\includegraphics[width=0.22\textwidth]
    		{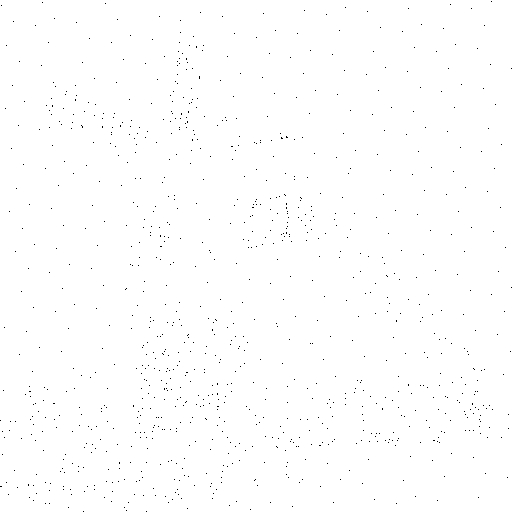} \\
    	
    	\multirow{1}{*}[7ex]{\rotatebox[origin=c]{90}{$\partial_x$}} &
    	~ &
    	\includegraphics[width=0.22\textwidth]
    		{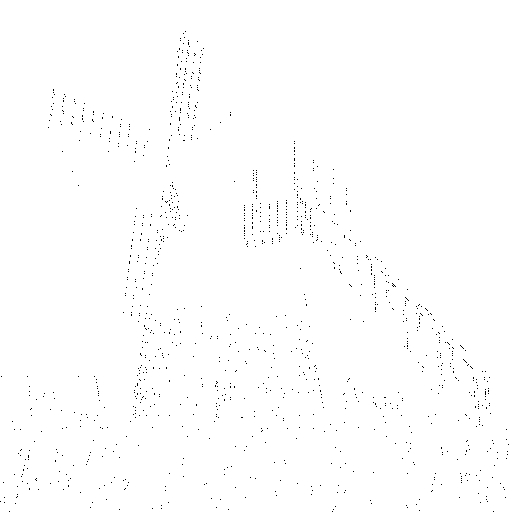} &
    	\includegraphics[width=0.22\textwidth]
    		{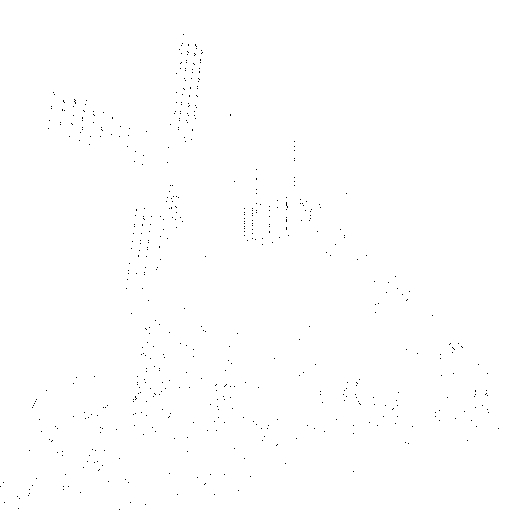} &
    	\includegraphics[width=0.22\textwidth]
    		{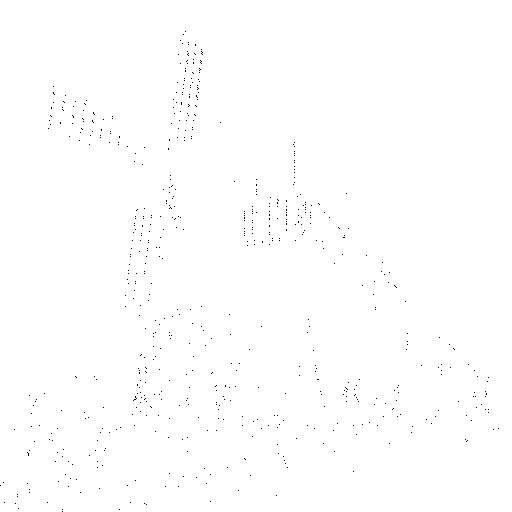} \\
    	
    	\multirow{1}{*}[7ex]{\rotatebox[origin=c]{90}{$\partial_y$}} &
    	~ &
    	\includegraphics[width=0.22\textwidth]
    		{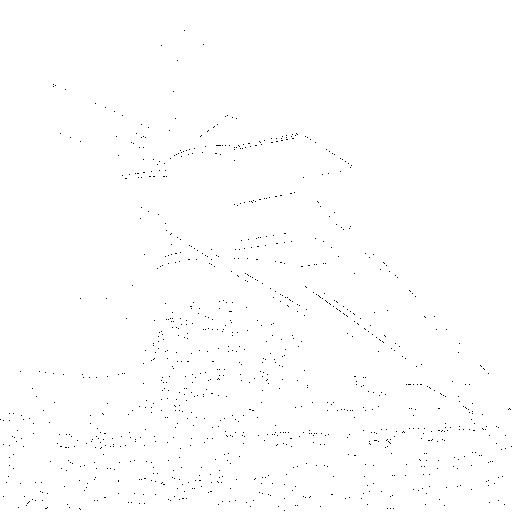} &
    	\includegraphics[width=0.22\textwidth]
    		{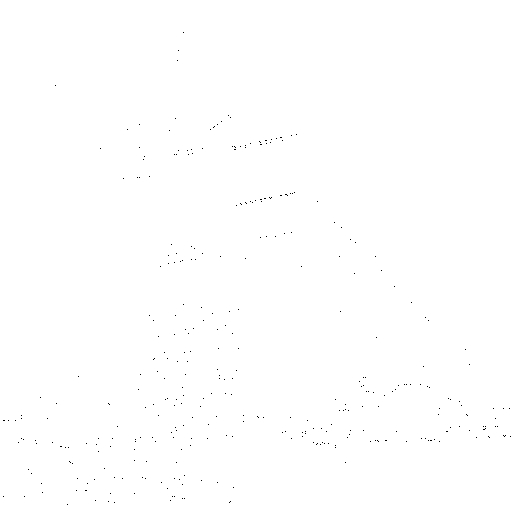} &
    	\includegraphics[width=0.22\textwidth]
    		{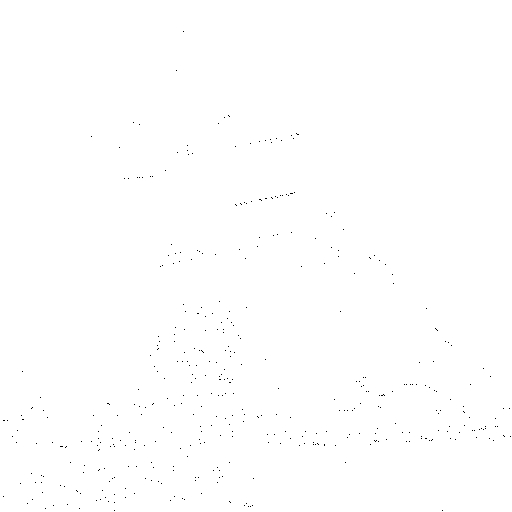} \\
    	
    	\multirow{1}{*}[9ex]{\rotatebox[origin=c]{90}{$3 \times 3$}} &
    	~ &
    	~ &
    	\includegraphics[width=0.22\textwidth]
    		{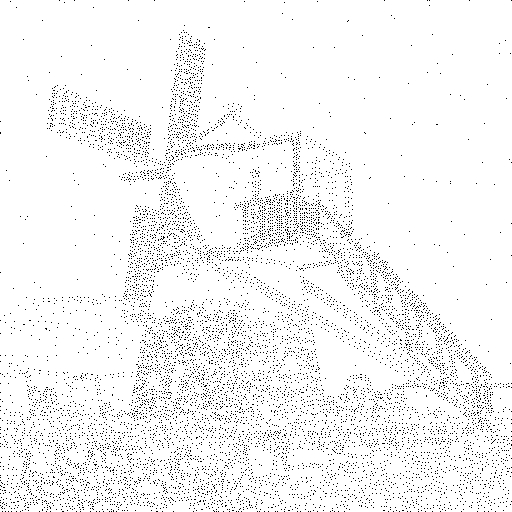} &
    	\includegraphics[width=0.22\textwidth]
    		{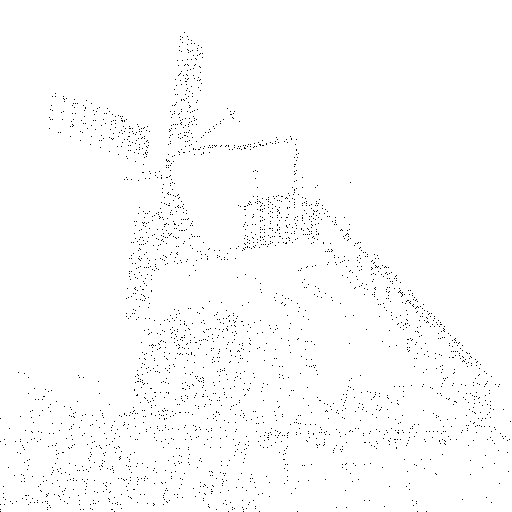} \\
    	
    	\multirow{1}{*}[9ex]{\rotatebox[origin=c]{90}{$5 \times 5$}} &
    	~ &
    	~ &
    	~ &
    	\includegraphics[width=0.22\textwidth]
    		{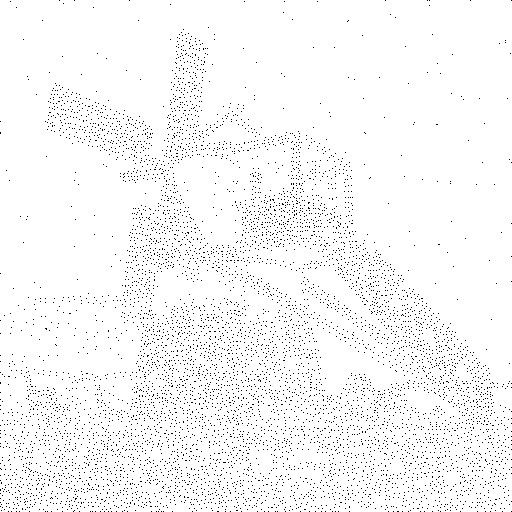} \\
    	
    	~ &
            MSE:~$185.13$ &
            MSE:~$162.10$ &
    	MSE:~$81.18$ &
    	MSE:~$\bm{78.95}$ \\
    	
    \end{tabularx}
    }
    \caption{Harmonic inpainting results for a combined density of $5\%$ 
    for \emph{windmill} produced with 30 densification iterations.}
    \label{fig:exp_qualitative_windmill}
\end{figure}

Inpaintings along with masks for an increasing 
variety of considered features are presented in
\Cref{fig:exp_qualitative_elpaso} and \Cref{fig:exp_qualitative_windmill}.
The figures demonstrate the effect of increasing the number of features 
while keeping the same total mask density -- the mean squared error (MSE) 
generally decreases in a monotone way. This testifies to the robustness of
our optimization and also illustrates the viability of the chosen features. 

\paragraph{Improved Densification}
We also improve upon the results from our conference publication~\cite{JCW23}. 
There the five features' MSEs for \emph{elpaso} and \emph{windmill} at 
5\% were respectively $23.25$ and $157.30$. With our new densification 
we are able to achieve MSEs of $15.27$ and $78.95$.
This is mainly due to the improvements of our densification algorithm, 
and partially due to replacing the $2\times 2$ and 
$16\times 16$ averages with the $3\times 3$ and $5\times 5$ binomial 
kernels.

\subsection{EED Inpainting with Linear Features}
\label{sec:exp_EED}

We use edge-enhancing diffusion (EED) for nonlinear inpainting 
because it has demonstrated very strong performance for sparse image 
inpainting with low- to mid-frequency content~\cite{GWWB08}.

\begin{figure}
    \begin{tabular}{ccc}
    \includegraphics[width=0.3\textwidth]{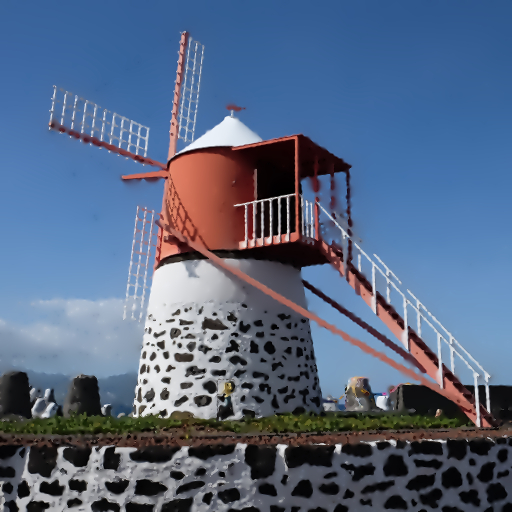} &
    \includegraphics[width=0.3\textwidth]{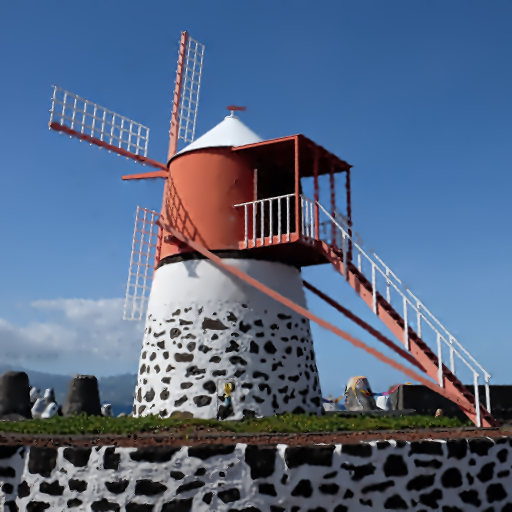} &
    \includegraphics[width=0.3\textwidth]{resources/windmill-512.jpg} 
    \\
    color values only & all five features & original
    \end{tabular}
    \caption{EED inpainting with linear features, 5\% mask. 
    Note the much sharper result and better reconstructed 
    edges.}
    \label{fig:exp_visual_widmill_EED3}
\end{figure}

\begin{figure}
\begin{center}
    \begin{tabular}{ccc}
    \includegraphics[width=0.3\textwidth]{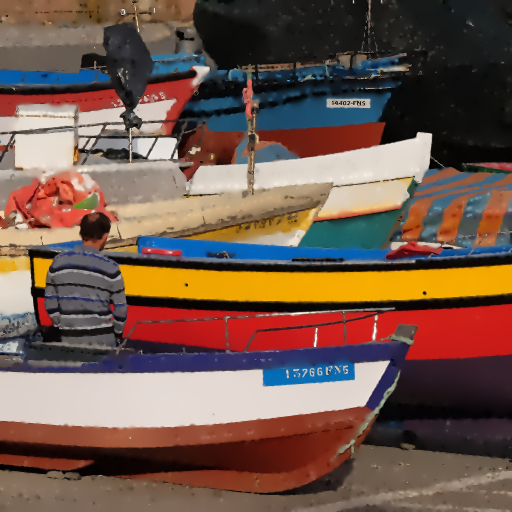}
    \hspace{-2.55mm} & \hspace{-2.55mm}
    \includegraphics[width=0.3\textwidth]{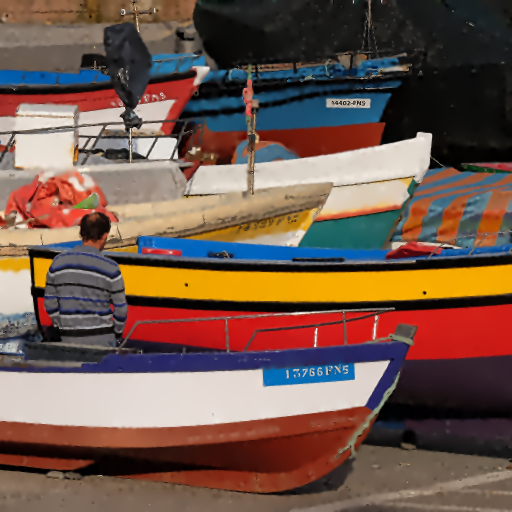}
    \hspace{-2.55mm} & \hspace{-2.55mm}
    \includegraphics[width=0.3\textwidth]{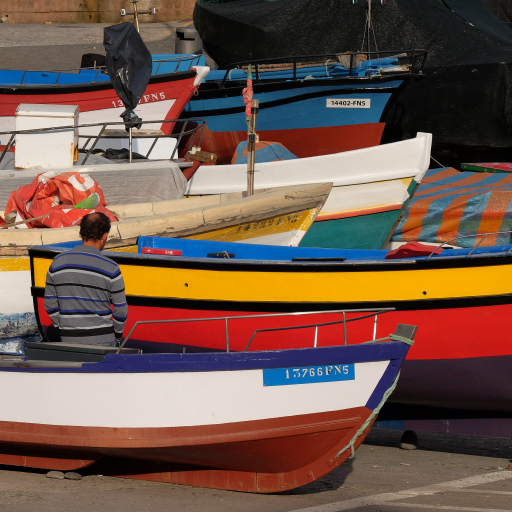}  
    \\
    \includegraphics[width=0.3\textwidth]{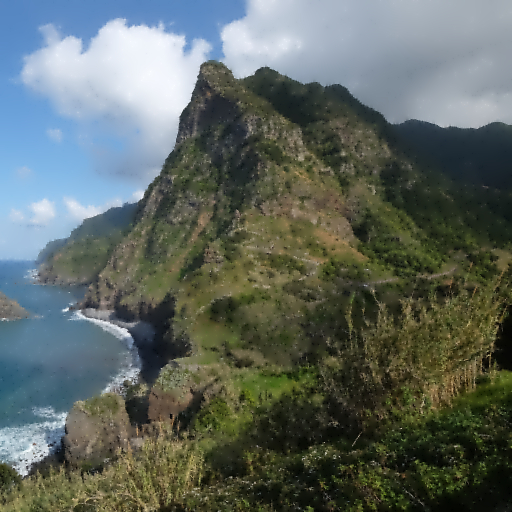}
    \hspace{-2.55mm} & \hspace{-2.55mm}
    \includegraphics[width=0.3\textwidth]{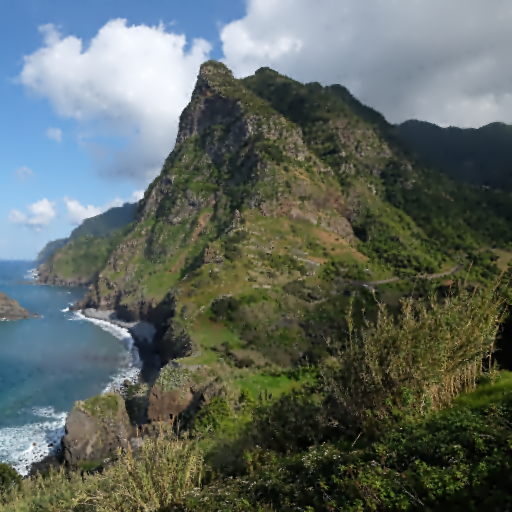}
    \hspace{-2.55mm} & \hspace{-2.55mm}
    \includegraphics[width=0.3\textwidth]{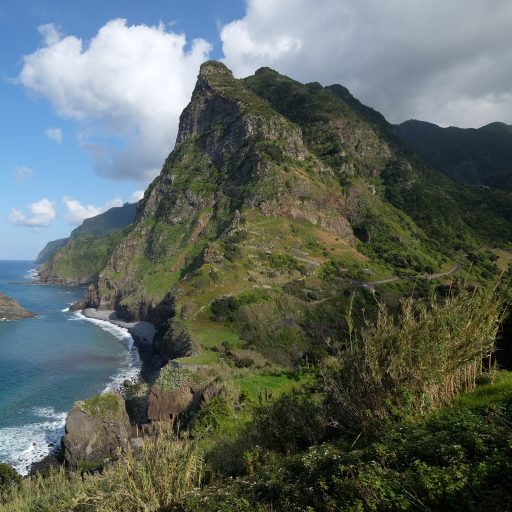}  
    \\
    \includegraphics[width=0.3\textwidth]{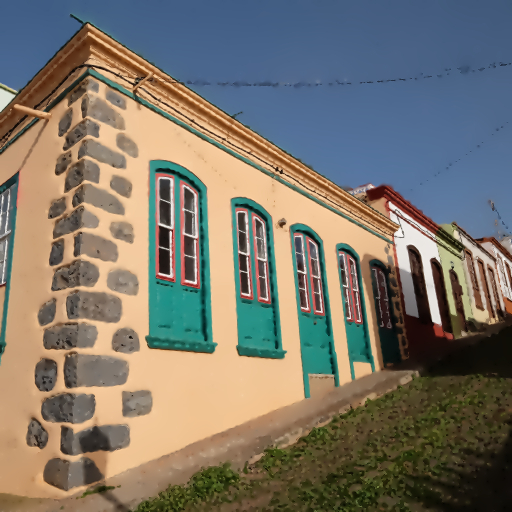}
    \hspace{-2.55mm} & \hspace{-2.55mm}
    \includegraphics[width=0.3\textwidth]{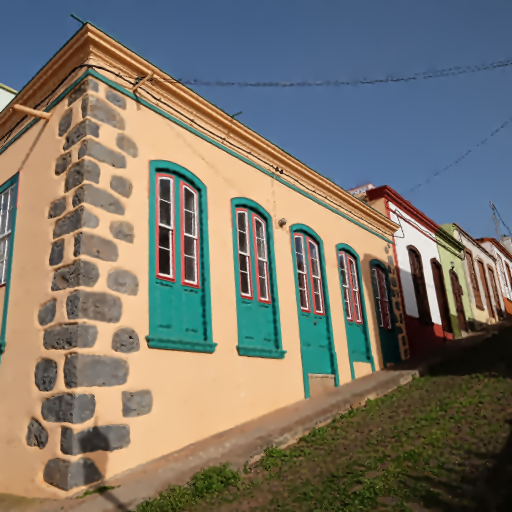}
    \hspace{-2.55mm} & \hspace{-2.55mm}
    \includegraphics[width=0.3\textwidth]{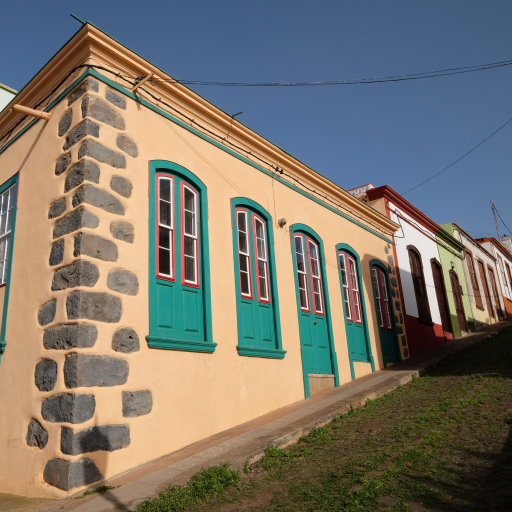}  
    \\
    \includegraphics[width=0.3\textwidth]{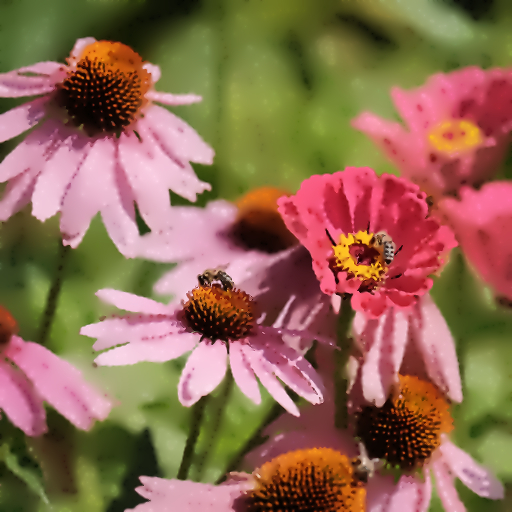}
    \hspace{-2.55mm} & \hspace{-2.55mm}
    \includegraphics[width=0.3\textwidth]{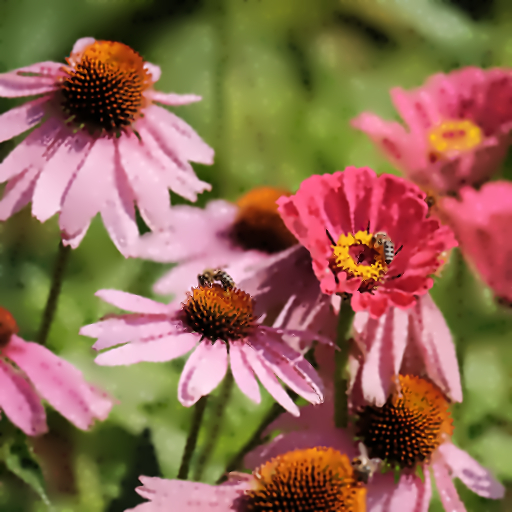}
    \hspace{-2.55mm} & \hspace{-2.55mm}
    \includegraphics[width=0.3\textwidth]{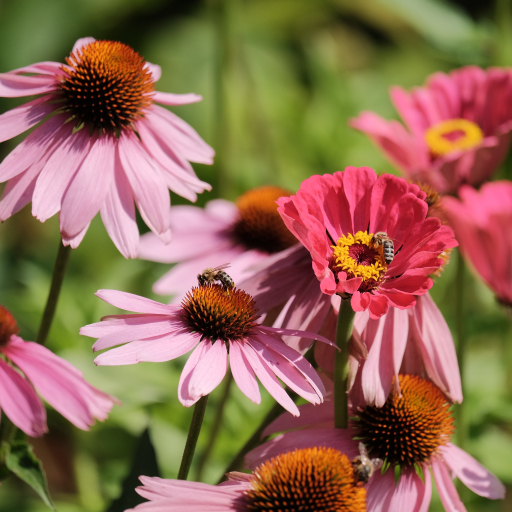}  
    \\
    \end{tabular}
\end{center}
    \caption{EED inpainting. Left: color values only; 
    middle: all five feature types; right: original. Photos 
    by Joachim Weickert -- top to bottom: 
    \textit{boats, madeira, garafia, flowers}.}
    \label{fig:exp_collected_EED}
\end{figure}

\paragraph{Visual Comparison}
\Cref{fig:exp_visual_widmill_EED3} illustrates the visual difference 
between EED inpainting with only color interpolation and 
with the five features. The latter is less blurry and wavy and has closed 
some image edges better. 
Additional visual comparisons can be found in 
\cref{fig:exp_collected_EED}. 

\paragraph{Mean Squared Error}
\Cref{tbl:MSE_table_EED} shows the mean squared errors (MSEs) for all twelve 
images using an increasing set of feature types $\{\bm{\phi}_k\}_{k=1}^{q}$, 
while keeping the total mask density constant. Note that the original 
images have range $\{0,1,\ldots,255\}$. We achieve up to $40$\% 
reduction in the MSE. For the EED inpainting setting we get a strong 
\textbf{average PSNR improvement of $\bm{1.82}$ dB} when comparing 
the setting of all five features ($q=5$) and only one feature ($q=1$). 

Similar to the harmonic inpainting case, the most impactful feature -- 
aside from the color interpolation -- is the binomial kernel feature 
$\bm{\phi}_4$. In contrast, the derivative features 
$\bm{\phi}_2$ and $\bm{\phi}_3$ have only a minor effect. This is because EED 
can inherently reconstruct edges from color values alone, making 
gradient-like features somewhat redundant.
This highlights an important principle in feature selection: \textbf{Features 
should be chosen to complement the specific weaknesses of the 
inpainting operator}. 
In the harmonic inpainting case the $3\times 3$ and $5\times 5$ stencils 
greatly help with suppressing the logarithmic singularities, 
while the derivative features help to bring some directional 
edge data which harmonic inpainting lacks due to its isotropic 
behavior. The latter is not the case for EED which can benefit from its
strong anisotropy.
%
%
\revhl{
\paragraph{Density Scaling in Terms of MSE}
For the image \textit{trui}, we additionally study how the mask 
density varies with a prescribed target MSE. As shown in 
\Cref{tbl:trui_eed_scaling}, doubling the target MSE reduces the 
required mask density to approximately two thirds of its previous 
value. At the same time, using five feature families instead of 
a single feature reduces the density required to attain a given 
reconstruction quality by approximately $44\%$--$55\%$, with 
the largest reductions occurring at the lowest densities.}
%
%
%
%
\begin{table}
\centering
\begin{tabular}{|l|c|c|c|c|c|c|}
\hline
$q$ & \textit{boats} 7\% & \textit{elpaso} 3\% & \textit{shed} 15\% & 
\textit{generator} 3\% & \textit{raindeer} 3\% & \textit{quai} 3\% \\
\hline
1 & 53.60 & 25.51 & 109.65 & 131.34 & 59.78 & 52.33 \\
\hline
3 & 51.80 & 24.07 &  93.64 & 128.23 & 58.17 & 51.01 \\
\hline
4 & 35.38 & 17.98 &  81.24 &  97.00 & 35.15 & 40.21\\
\hline
5 & \textbf{33.07} & \textbf{17.04} &  \textbf{79.02} &  \textbf{93.19} 
& \textbf{33.17} & \textbf{37.52} \\ 
\hline 
\hline
$q$ & \textit{mirror 3\%} & \textit{madeira 8\%} & \textit{garafia 7\%} & 
\textit{flowers 3\%} & \textit{crab 8\%} & \textit{windmill 5\%} \\
\hline
1 & 31.94 & 75.92 & 67.26 & 112.15 & 328.00 & 91.00 \\
\hline 
3 & 30.42 & 69.49 & 67.09 & 111.95 & 297.27 & 85.92 \\
\hline 
4 & 19.87 & 58.88 & 44.22 &  74.29 & 241.04 & 61.13 \\
\hline 
5 & \textbf{19.11} & \textbf{57.99} & \textbf{42.26} &  \textbf{67.84} & \textbf{236.42} & \textbf{56.35} \\
\hline
\end{tabular}
\caption{MSEs for EED inpainting  with an increasing number of 
feature families.}
\label{tbl:MSE_table_EED}
\end{table}
%
%
\begin{table}
\centering
\begin{tabular}{|l|c|c|c|c|}
\hline
target MSE & 20 & 40 & 80 & 160 \\
\hline
1 feature
& 5.62\% 
& 3.80\% 
& 2.63\% 
& 1.80\%  \\
\hline
5 features
& 3.14\%
& 1.88\%
& 1.23\%
& 0.81\% \\
\hline
\end{tabular}
\caption{\revhl{Mask density in percent required to reach the specified
target MSE for EED inpainting of \textit{trui}. 
 Doubling the target MSE reduces the required
mask density by approximately a factor of $2/3$. 
Using five features reduces the required density by 
$44\%$--$55\%$.}}
\label{tbl:trui_eed_scaling}
\end{table}
%

\paragraph{Inpaintings and Masks}

\begin{figure}
\center
    {
    \setlength{\tabcolsep}{1pt}
    \begin{tabularx}{\textwidth}{>{\columncolor{gray!10}}c >{\columncolor{gray!10}}c 
    >{\columncolor{gray!10}}c >{\columncolor{gray!10}}c >{\columncolor{gray!10}}c}
    	\multirow{1}{*}[13ex]{\rotatebox[origin=c]{90}{inpainting}} &
    	\includegraphics[width=0.22\textwidth]
    		{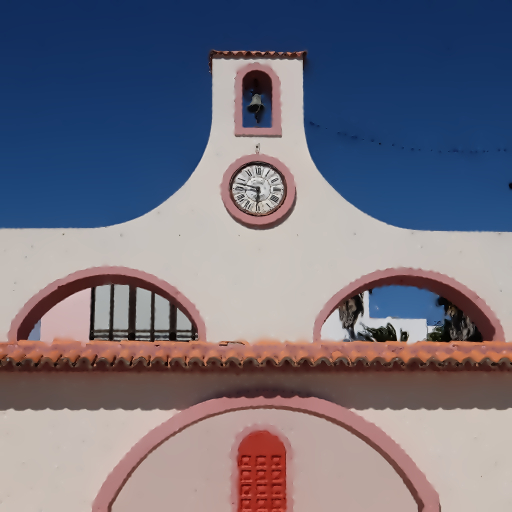} &
    	\includegraphics[width=0.22\textwidth]
    		{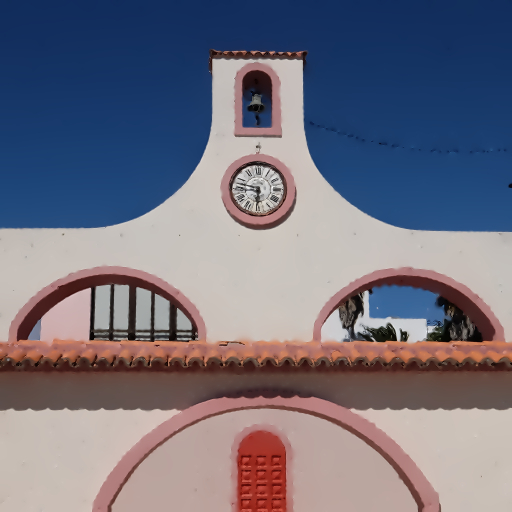} &
    	\includegraphics[width=0.22\textwidth]
    		{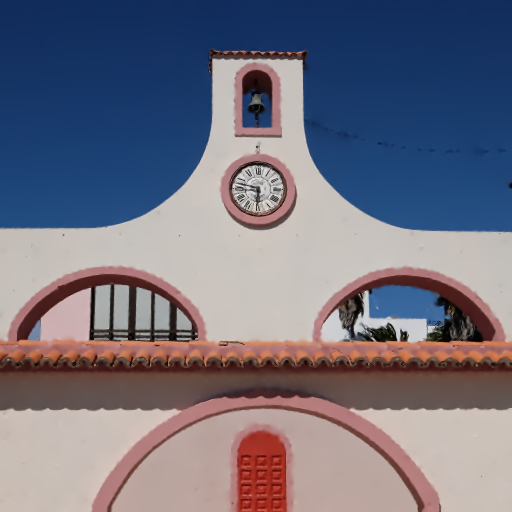} &
    	\includegraphics[width=0.22\textwidth]
    		{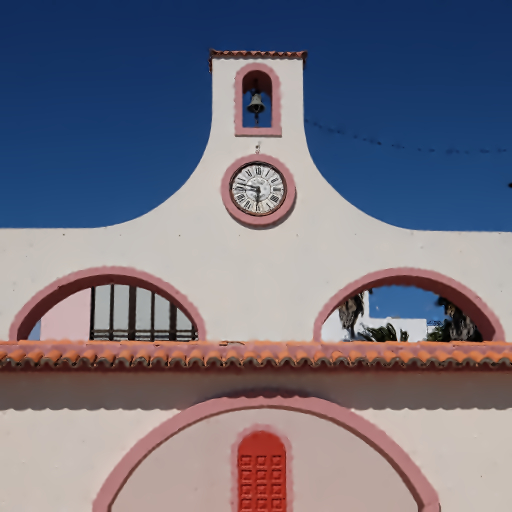} \\
    	
    	\multirow{1}{*}[8ex]{\rotatebox[origin=c]{90}{color}} &
    	\includegraphics[width=0.22\textwidth]
    		{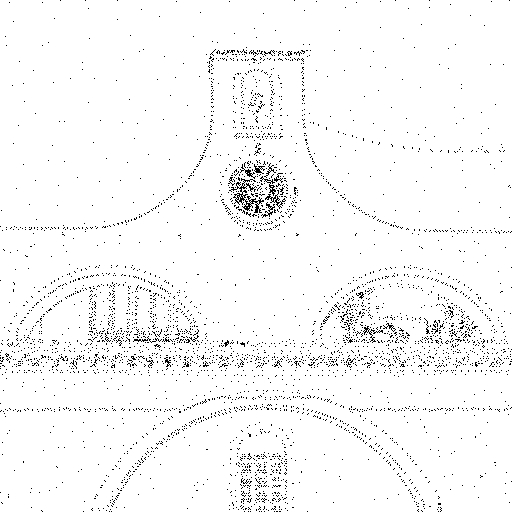} &
    	\includegraphics[width=0.22\textwidth]
    		{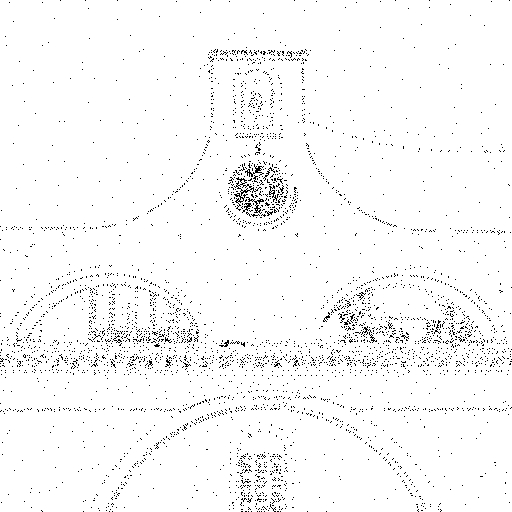} &
    	\includegraphics[width=0.22\textwidth]
    		{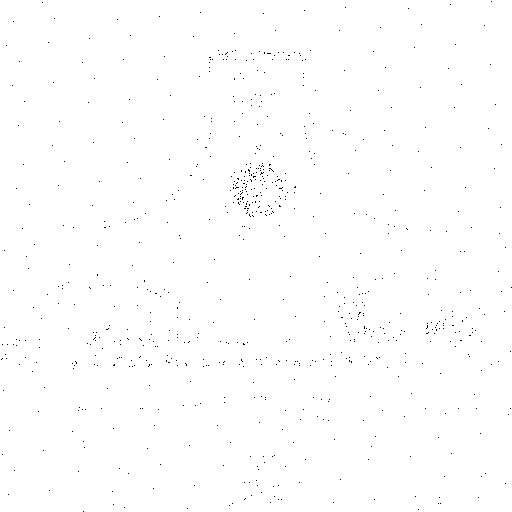} &
    	\includegraphics[width=0.22\textwidth]
    		{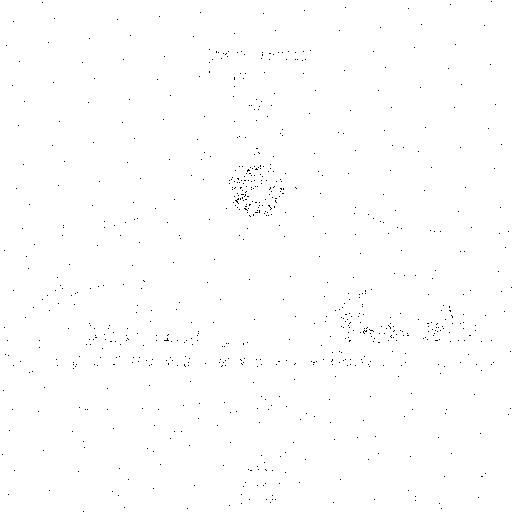} \\
    	
    	\multirow{1}{*}[7ex]{\rotatebox[origin=c]{90}{$\partial_x$}} &
    	~ &
    	\includegraphics[width=0.22\textwidth]
    		{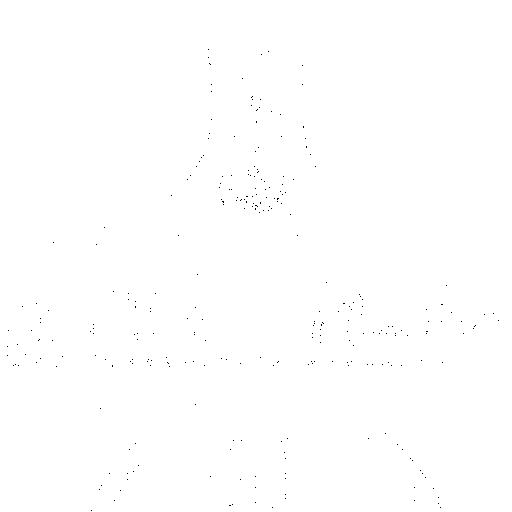} &
    	\includegraphics[width=0.22\textwidth]
    		{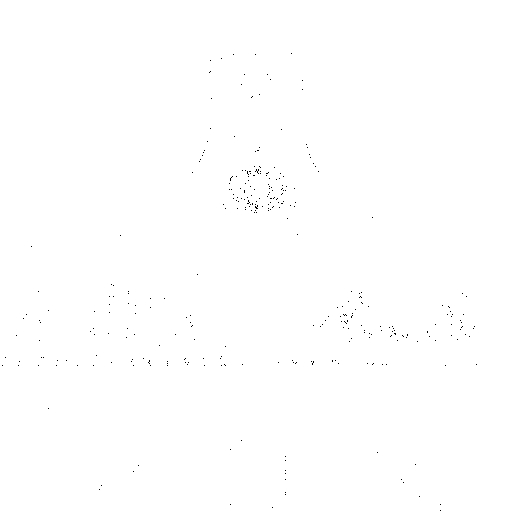} &
    	\includegraphics[width=0.22\textwidth]
    		{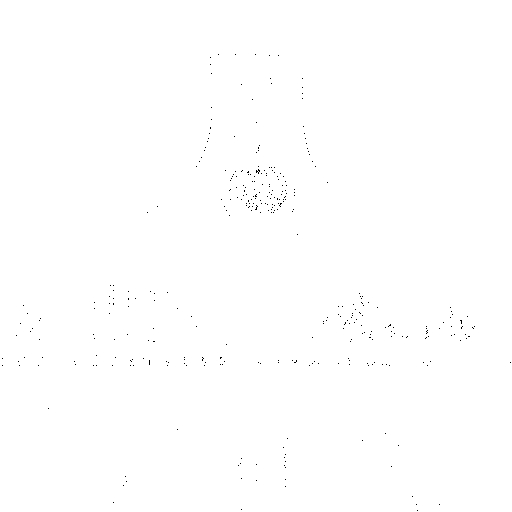} \\
    	
    	\multirow{1}{*}[7ex]{\rotatebox[origin=c]{90}{$\partial_y$}} &
    	~ &
    	\includegraphics[width=0.22\textwidth]
    		{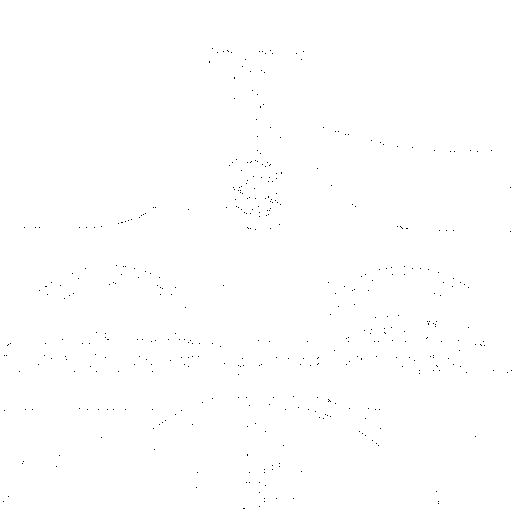} &
    	\includegraphics[width=0.22\textwidth]
    		{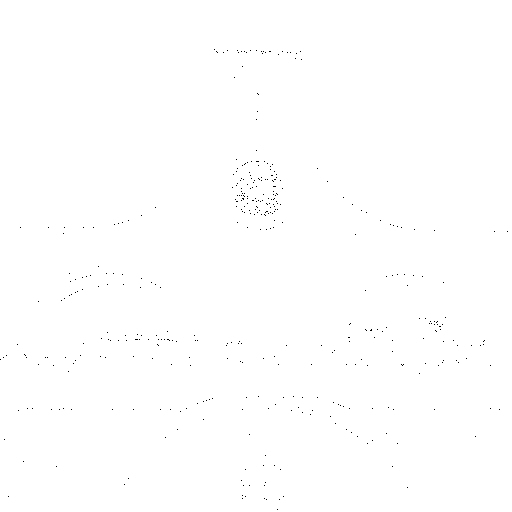} &
    	\includegraphics[width=0.22\textwidth]
    		{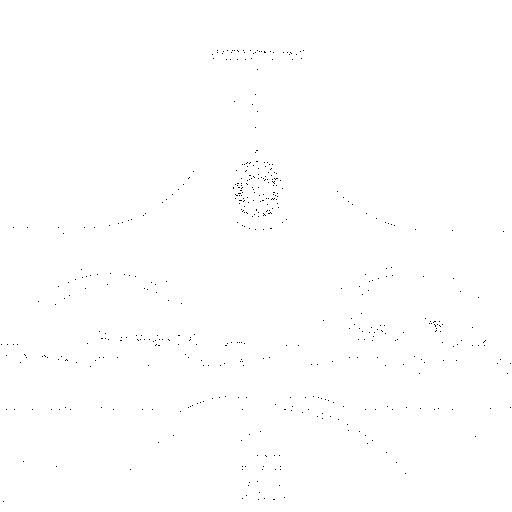}  \\
    	
    	\multirow{1}{*}[9ex]{\rotatebox[origin=c]{90}{$3 \times 3$}} &
    	~ &
    	~ &
    	\includegraphics[width=0.22\textwidth]
    		{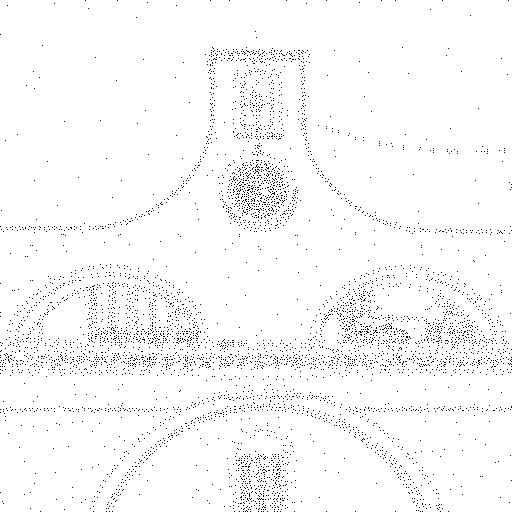} &
    	\includegraphics[width=0.22\textwidth]
    		{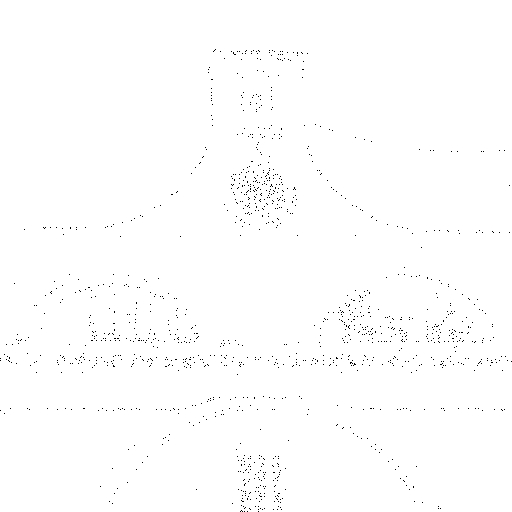} \\
    	
    	\multirow{1}{*}[9ex]{\rotatebox[origin=c]{90}{$5 \times 5$}} &
    	~ &
    	~ &
    	~ &
    	\includegraphics[width=0.22\textwidth]
    		{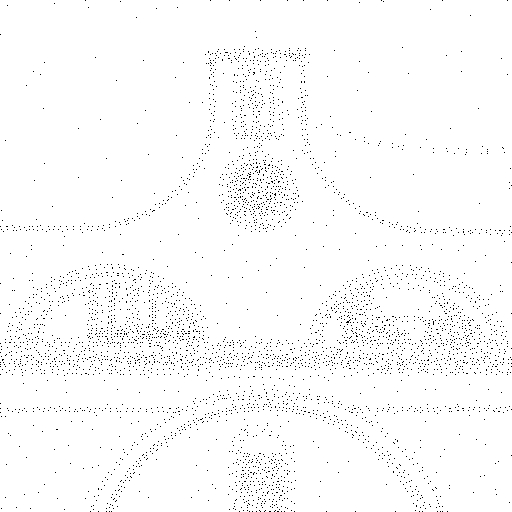} \\
    	
    	~ &
    	MSE:~$25.21$ &
    	MSE:~$24.07$ &
    	MSE:~$17.98$ &
    	MSE:~$\bm{17.04}$ \\
    \end{tabularx}
    }
    \caption{EED inpainting results for a combined density of $3\%$ 
    for \emph{elpaso} produced with 30 densification iterations}
    \label{fig:exp_qualitative_elpaso_EED3}
\end{figure}

\begin{figure}
\center
    {
    \setlength{\tabcolsep}{1pt}
    \begin{tabularx}{\textwidth}{>{\columncolor{gray!10}}c >{\columncolor{gray!10}}c 
    >{\columncolor{gray!10}}c >{\columncolor{gray!10}}c >{\columncolor{gray!10}}c}
    	\multirow{1}{*}[13ex]{\rotatebox[origin=c]{90}{inpainting}} &
    	\includegraphics[width=0.22\textwidth]
    		{resources/EED3/windmill-512_c3_d0_05_EED3_ni30_ns1.jpg} &
    	\includegraphics[width=0.22\textwidth]
    		{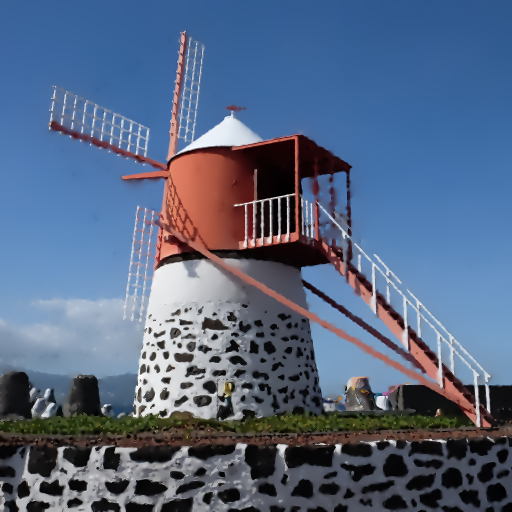} &
    	\includegraphics[width=0.22\textwidth]
    		{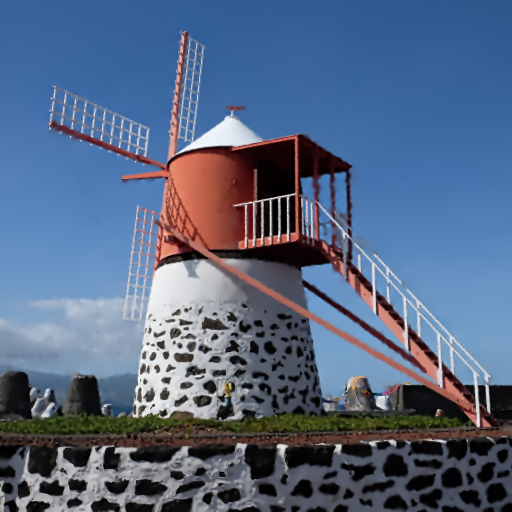} &
    	\includegraphics[width=0.22\textwidth]
    		{resources/EED3/windmill-512_c3_d0_05_EED3_ni30_ns5.jpg} \\
    	
    	\multirow{1}{*}[8ex]{\rotatebox[origin=c]{90}{color}} &
    	\includegraphics[width=0.22\textwidth]
    		{resources/harmonic/windmill/windmill-512_c3_d0_05_EED0_ni30_ns1_fi1.jpg} &
    	\includegraphics[width=0.22\textwidth]
    		{resources/harmonic/windmill/windmill-512_c3_d0_05_EED0_ni30_ns3_fi1.jpg} &
    	\includegraphics[width=0.22\textwidth]
    		{resources/harmonic/windmill/windmill-512_c3_d0_05_EED0_ni30_ns4_fi1.jpg} &
    	\includegraphics[width=0.22\textwidth]
    		{resources/harmonic/windmill/windmill-512_c3_d0_05_EED0_ni30_ns5_fi1.jpg} \\
    	
    	\multirow{1}{*}[7ex]{\rotatebox[origin=c]{90}{$\partial_x$}} &
    	~ &
    	\includegraphics[width=0.22\textwidth]
    		{resources/harmonic/windmill/windmill-512_c3_d0_05_EED0_ni30_ns3_fi2.jpg} &
    	\includegraphics[width=0.22\textwidth]
    		{resources/harmonic/windmill/windmill-512_c3_d0_05_EED0_ni30_ns4_fi2.jpg} &
    	\includegraphics[width=0.22\textwidth]
    		{resources/harmonic/windmill/windmill-512_c3_d0_05_EED0_ni30_ns5_fi2.jpg} \\
    	
    	\multirow{1}{*}[7ex]{\rotatebox[origin=c]{90}{$\partial_y$}} &
    	~ &
    	\includegraphics[width=0.22\textwidth]
    		{resources/harmonic/windmill/windmill-512_c3_d0_05_EED0_ni30_ns3_fi3.jpg} &
    	\includegraphics[width=0.22\textwidth]
    		{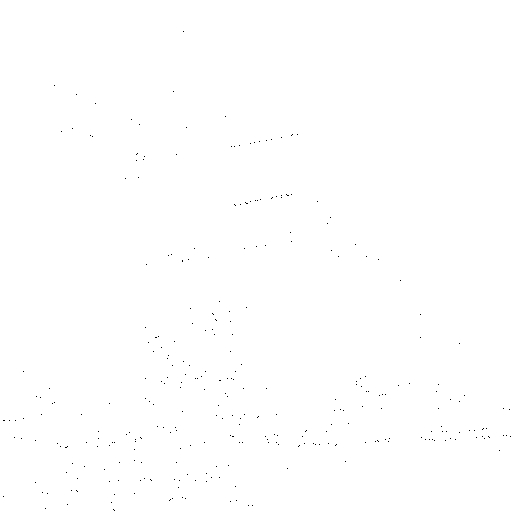} &
    	\includegraphics[width=0.22\textwidth]
    		{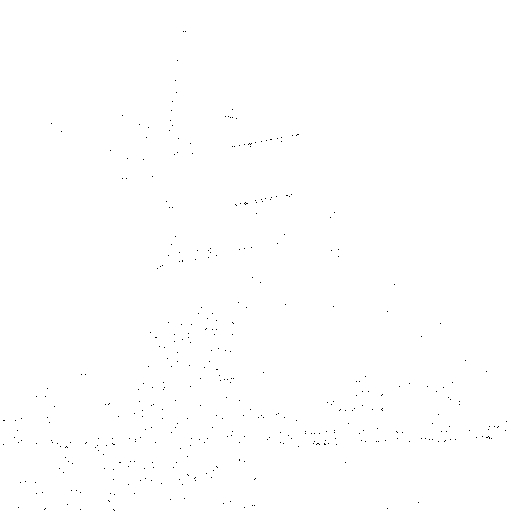} \\
    	
    	\multirow{1}{*}[9ex]{\rotatebox[origin=c]{90}{$3 \times 3$}} &
    	~ &
    	~ &
    	\includegraphics[width=0.22\textwidth]
    		{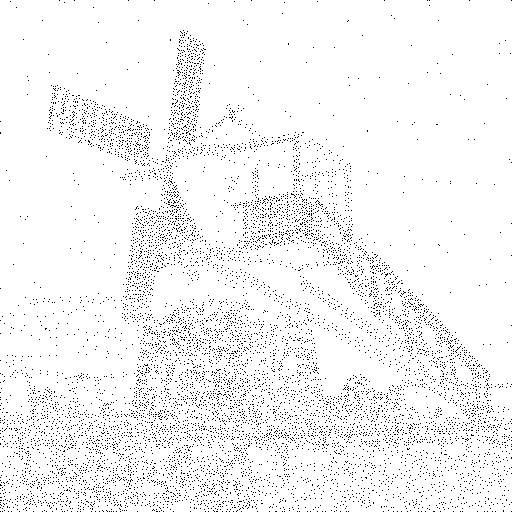} &
    	\includegraphics[width=0.22\textwidth]
    		{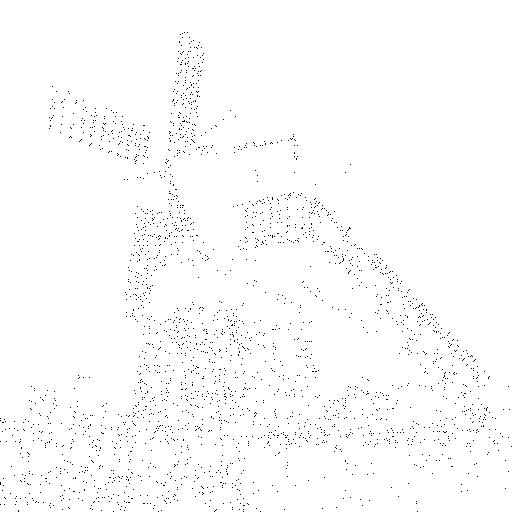} \\
    	
    	\multirow{1}{*}[9ex]{\rotatebox[origin=c]{90}{$5 \times 5$}} &
    	~ &
    	~ &
    	~ &
    	\includegraphics[width=0.22\textwidth]
    		{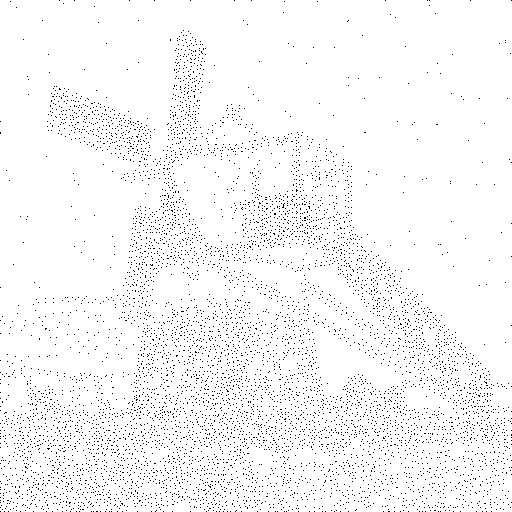} \\
    	
    	~ &
            MSE:~$91.00$ &
            MSE:~$88.25$ &
    	MSE:~$61.13$ &
    	MSE:~$\bm{56.35}$ \\
    	
    \end{tabularx}
    }
    \caption{EED inpainting results for a combined density of $5\%$ 
    for \emph{windmill} produced with 30 densification iterations.}
    \label{fig:exp_qualitative_windmill_EED3}
\end{figure}

\Cref{fig:exp_qualitative_elpaso_EED3} and \Cref{fig:exp_qualitative_windmill_EED3} 
show inpainted images and their corresponding masks for an increasing 
number of feature types. These examples illustrate the effect of adding 
more features while keeping the total mask density fixed—resulting in a 
generally monotonic decrease in mean squared error (MSE). The same images 
are chosen as in the harmonic inpainting case in order to facilitate 
comparisons.


\subsection{Inpainting with Nonlinear Features}
\label{exp:nlin_nlin}

\begin{figure}
    \begin{tabular}{ccc}
    \includegraphics[width=0.3\textwidth]{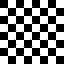} &
    \includegraphics[width=0.3\textwidth]{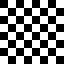} &
    \includegraphics[width=0.3\textwidth]{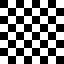} 
    \\
    lin.\ constr., MSE: $43.65$ & 
    nonlin.\ constr., MSE: $1.53$ &
    original
    \end{tabular}
    \caption{Checkered texture reconstructed with $30$ 
    densification iterations at $40\%$ density: 
    pointwise (left) vs pointwise and 
    the nonlinear affine-invariant curvature features (middle).}
    \label{fig:exp_visual_widmill_EED0_nlin}
\end{figure}

In the context of nonlinear features, we compare two setups: harmonic 
inpainting using the classical pointwise interpolation feature $\phi_1$, and 
harmonic inpainting using $\phi_1$ and the affine-invariant curvature 
feature $\tilde{\phi_2}$. To make the effect of 
the curvature feature obvious, we consider 
an image that has sufficiently many corners -- in this case a $64\times 64$ 
checkered texture with $8 \times 8$ tiles.
Our results are shown in \cref{fig:exp_visual_widmill_EED0_nlin}, demonstrating 
a large MSE improvement. For natural images the affine-invariant 
curvature feature provides very minor benefits, because the number of important 
corners are generally many times fewer than smoother structures.
Identifying nonlinear features that are more suitable for natural images 
remains an open direction for future research.

%
%

\section{Conclusions and Future Work}
\label{sec:conclusions}

\revhl{Building on our previous work in the linear setting~\cite{JCW23}, 
we have presented a unified framework for feature-based inpainting 
with equality constraints and extended it to nonlinear inpainting
 operators and nonlinear constraints.} The central insight is that it is 
sufficient to replace the mask matrix 
$\bm{C}$ with the orthogonal projection matrix 
$\bm{P} = \bm{J}_{\bm{b}}^+\bm{J}_{\bm{b}}$, which rejects the kernel 
of the constraints' Jacobian. This formulation aligns with optimization-based 
approaches and remains applicable even when the inpainting operator 
is not derived from an energy functional.
Our framework also highlights an important conceptual point: Prescribing a 
set of features inherently alters the inpainting operator, e.g., 
logarithmic singularities inherent to harmonic inpainting can disappear. 
This is explained by the fact that the inpainting problem being solved 
involves the projection of the inpainting operator on the kernel of 
the linearized constraints $\bm{I}-\bm{P}$. 

We have also extended our spatial optimization method 
from~\cite{JCW23} to support 
nonlinear inpainting operators and nonlinear constraints. The 
significant quality gains achieved simply by incorporating 
additional features emphasize that improving the modeling of 
the reconstruction process can be as impactful -- if not more so -- 
as optimizing the data selection itself.

Our experiments with edge-enhancing diffusion (EED) 
inpainting revealed that derivative features contribute little 
to its reconstruction quality. This is consistent with the fact that 
EED already excels at edge completion. The takeaway is that 
features should be chosen to complement the specific limitations
of the inpainting operator, targeting structures it cannot 
reconstruct effectively on its own. This idea is similar in nature 
to the results about the complementarity of the data 
term and the smoothness term in optic flow~\cite{ZBW11}.

Finally, we emphasize that the consistent reduction in mean squared error (MSE), reaching up to 60\% in our feature inpainting experiments, highlights the practical relevance of our 
approach, particularly for applications such as inpainting-based 
image compression. 
Notably, the features contributing most to this reduction -- local low-pass integrals -- are also the least sensitive to quantization, further underscoring their applicability to the compression setting. We have designed our algorithms with practical applications in mind: Our densification approach automatically determines feature ratios and their distribution without requiring external input. Moreover, it offers a straightforward mechanism to balance runtime and reconstruction quality via a single intuitive parameter -- 
the number of densification iterations.

Although our framework is very general -- supporting a wide range of 
features and inpainting strategies -- we would like to emphasize 
that there are still various ways to extend it.
One such avenue involves incorporating inequality constraints. 
Another possible direction is identifying novel nonlinear 
features that offer further quality improvements. 
Additionally, (sub-)gradient descent based data selection strategies 
present a natural post-processing step following our densification, 
offering the potential for even better reconstructions.  
\revhl{
From a practical perspective, a natural next step is the 
development of computationally efficient implementations 
and embedding the proposed framework into 
a complete codec with quantization, entropy coding, and
 rate-distortion optimization, thereby enabling a comprehensive 
 assessment of its compression performance.}
Finally, machine learning techniques could be investigated 
as a means to accelerate both the inpainting and data 
selection processes.


%


\bibliographystyle{siamplain}
\bibliography{myrefs,additional_refs}
	
\end{document}